\newlength\gsfigwidth
\newlength\figwidth
\documentclass[aps,prb,twocolumn,floatfix,amsmath,amssymb,showpacs,showkeys]{revtex4}

\newcommand{\be}{\begin{equation}}
\newcommand{\ee}{\end{equation}}
\newcommand{\ba}{\begin{align}}
\newcommand{\ea}{\end{align}}
\newcommand{\bn}{\begin{eqnarray}}
\newcommand{\en}{\end{eqnarray}}

\usepackage{graphicx}

\begin{document}

\preprint{KSU-UFV-Wysin et al.}

\title{Order and thermalized dynamics in Heisenberg-like square and Kagom\'e spin ices}
\author{G.\ M.\  Wysin }
\email{wysin@phys.ksu.edu}
\homepage{http://www.phys.ksu.edu/personal/wysin}
\affiliation{Department of Physics, Kansas State University, Manhattan, KS 66506-2601}

\author{A. R. Pereira}
\email{apereira@ufv.br.}
\homepage{https://sites.google.com/site/quantumafra/home}
\affiliation{Departamento de F\'{i}sica, Universidade Federal de Vi\c{c}osa, 36570-000 - Vi\c{c}osa - Minas Gerais - Brazil.}

\author{W. A. Moura-Melo}
\email{winder@ufv.br}
\homepage{https://sites.google.com/site/wamouramelo/}
\affiliation{Departamento de F\'{i}sica, Universidade Federal de Vi\c{c}osa, 36570-000 - Vi\c{c}osa - Minas Gerais - Brazil.}

\author{C.I.L. de Araujo}
\email{dearaujo@ufv.br}
\affiliation{Departamento de F\'{i}sica, Universidade Federal de Vi\c{c}osa, 36570-000 - Vi\c{c}osa - Minas Gerais - Brazil.}

\date{\today}
\begin{abstract}
{Thermodynamic properties of a spin ice model on a Kagom\'e lattice are obtained from dynamic simulations
and compared with properties in square lattice spin ice.  The model assumes three-component 
Heisenberg-like dipoles of an array of planar magnetic islands situated on a Kagom\'e lattice.  
Ising variables are avoided. The island dipoles interact via long-range dipolar interactions and are 
restricted in their motion due to local shape anisotropies.  We define various order parameters and obtain 
them and thermodynamic properties from  the dynamics of the system via a Langevin equation, solved by the 
Heun algorithm.  Generally, a slow cooling from high to low temperature does not lead to a particular 
state of order, even for a set of coupling parameters that gives well thermalized states and dynamics.  
Some suggestions are proposed for the alleviation of the geometric frustration effects and for the 
generation of local order in the low temperature regime.}
\end{abstract}

\maketitle

\section{Introduction: Spin ice frustration and dynamics}
Recently, nanostructured arrays of magnetic materials known as artificial spin ices \cite{Wang2006,Moler2006,Mol2009,Ladak2010,Mengotti2011,Morgan2011}
have received a lot of attention due to their interesting characteristics: firstly, they are intentionally built (with the use of magnetic 
nanotechnology) to display geometrical frustration; secondly, they may support excitations that behave like magnetic 
monopoles \cite{Mol2009,Mol2010,Moler09,Silva13,Nascimento12} and consequently, they are expected to be used in future technologies 
(as magnetricity and magnetronics). Initially these systems were experimentally performed, in general, by permalloy films with tickness 
between 20-30 nm and were found in frozen (athermal states) because the energy barriers that separate their microstates are much higher than 
the thermal energy. Such difficulty could be an obstacle for their manipulation and applications and, therefore, some researchers started finding
manners to overcome this problem and several protocols for adjusting artificial spin ice systems to undergo thermal fluctuations were 
successfully fashioned \cite{Kapaklis2012,Porro2013,Farhan2013,Zhang2013}. Among them, Morgan et al.\cite{Morgan2011}, observed ordered 
states in the early stage of film deposition and it opened a path for investigations in systems with film thickness lower than 3nm \cite{balan1,balan2},
where the variation of energy barriers allows superparamagnetic flutuations at temperatures close to the ambient. Then, the ground 
states of artificial spin ices were obtained and the excited states were experimentally identified. 
Such experimental advances also allow a more direct connection with theoretical results about the thermodynamics \cite{Silva12,Budrikis2012,Wysin2013} 
of artificial spin ices, mainly to obtain further insights into fundamental phenomena like geometrical frustration and phase transition.

In general, the elongated magnetic islands that form these artificial systems are well described theoretically by a resulting magnetic moment (spin)
with an Ising behavior pointing out along the longest axis of the islands. Then, in principle, these systems do not present a true dynamics. 
However, as in the case of the thermodynamics cited in the earlier paragraph, we could also expect future developments in materials preparations
and a subsequent difference in the islands magnetic behavior. Independent of it, the study of the frustration phenomenon in its diverse versions
and possibilities is also important for its complete physical understanding. Thus, differently from most previous works, in this paper we 
theoretically investigate two-dimensional spin ices in which the total spin of the islands is not Ising-like. Indeed, here, it has a kind of
anisotropic Heisenberg behavior and a coherent rotation. This spin then is a vector with three components but it prefers to be in the plane of 
the array and to point out along the longest axis of the nanoislands
due to anisotropies included into our model. We focus our attention in the order and thermalized dynamics of these artificial materials with 
emphasis on two types of geometries: the square and Kagom\'{e} lattices. The work is organized as follows: in Section $II$, we define some 
useful order parameters for square and Kagom\'e systems. In Section $III$, the thermal ordering in spin ices are simulated and the results 
for square and Kagom\'e arrays are presented in Sections $IV$ and $V$ respectively. Finally, the discussions and conclusions are considered
in Section $VI$.      
\section{Distinguishing the order and ground states in spin ice}
A ground state of a spin ice should tend to satisfy the constraints of the local dipolar interactions, 
to avoid the high energy associated when like poles of the islands' dipoles are near each other.  
In spin ice on a square lattice, this leads to the ``two-in/two-out rule,''
where at any junction of four islands two of the dipoles point inward and two dipoles point
outward. This gives a net zero number of $+/-$ poles in a junction. If that takes place, that 
junction locally reduces its energy, and as well, does not contain a magnetic monopole charge.  
A similar rule applies for 3D spin ice in a tetragonal geometry.

The situation is somewhat different in Kagom\'e ice, because the islands form junctions
of three dipoles.  Hence, considering Ising-like dipoles, there will always be a magnetic
charge present in each junction.  If two dipoles point in (out) and one points out (in), there 
is a net negative (positive) monopole at that junction.  If all three point in (out), there is a 
triply charged negative (positive) magnetic pole at the junction.  In terms of energetics, the 
monopole charges will possess lesser energy than the triple charges.  However, only particular 
ordered arrangements of these poles will lead to ground states.  We will use the structure of 
the ground states to develop some parameters to indicate at the very least, the local ordering 
in Kagom\'e spin ice.

It is typical to develop theory based on Ising dipoles parallel/antiparallel to an island axis.
This restriction of movement is due to the shape anistropy (demagnetization effects) of the 
high-aspect ratio islands.  We avoid the Ising approach, because the island dipoles can have a 
certain freedom to tilt through small angles both out of the ($xy$) plane of the islands and within the 
plane of the islands.  Instead,  the islands are assumed to have a magnetic dipole of nearly
fixed magnitude $\mu$, but varying direction.  Thus, the $i$th island's dipole will be expressed
by a 3D vector $\vec\mu_{i}=\mu\hat{\mu}_{i}$.  This means we can just keep track of the directions
of the unit dipoles $\hat\mu_i$.  The tendency for the dipoles to align with the islands' long axes
is incorporated in the model with the use of local anisotropy interactions (a planar anisotropy
and a uniaxial anisotropy along the island axes \cite{Wysin2013,Wysin2012}.

\begin{figure}
\includegraphics[width=\gsfigwidth]{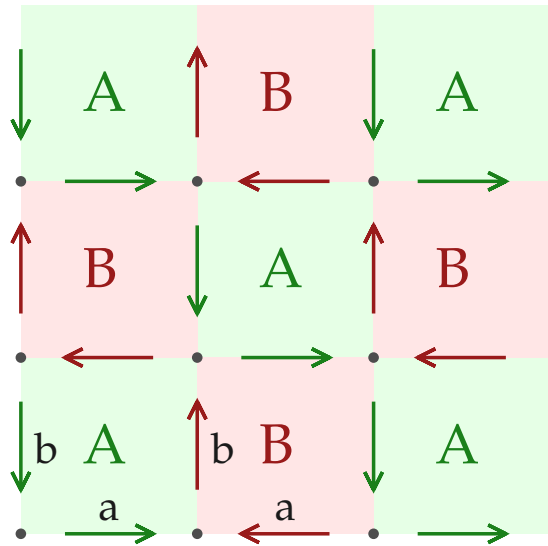}
\caption{\label{sqr-gs1} Square lattice spin ice in one of the ground states. A and B denote
the two sets of primitive cells. In each cell, a and b denote the magnetic basis sites.  A rotation $R$
of the B cells through 180$^{\circ}$ causes their a and b dipoles to align with those of the A cells.}
\end{figure}

\subsection{Square ice ground states and order}
It helps to look to square ice for some guidance about how to proceed in Kagom\'e ice.  In square 
ice pairs of islands belong to primitive cells of a square grid, see Figure \ref{sqr-gs1}. The primitive 
cells themselves belong to one of two sublattices, an $A$ sublattice and a $B$ sublattice.   The points 
$\vec{r}_k$ at one corner of each primitive cell act as monopole charge centers, located on 
a square lattice of lattice constant $a$.  The island pairs form an effective two-atom basis: 
one is centered at $\vec{r}_k+\tfrac{a}{2}\hat{x}$ and the other at $\vec{r}_k+\tfrac{a}{2}\hat{y}$.  
For the magnetic properties, we refer to them as magnetic positions, denoting these as $a$-sites and $b$-sites.  
The $a$-site ($b$-site) Ising-like dipoles would only point along $\pm\hat{x}$ ($\pm\hat{y}$), whereas,
the Heisenberg-like dipoles make deviations around these directions.  These directions are the islands' 
long axes. 

Figure  \ref{sqr-gs1} shows one of the two ground states for square ice, totally satisfying the two-in/two-out
rule.  The other ground state would be obtained by inverting all dipoles.  The primitive cell 
$k$ contains the two unit dipoles, $\hat\mu_{k,a}=\pm \hat{x}$ and $\hat\mu_{k,b}=\pm\hat{y}$.  
A ground state is 
\be
\hat\mu_{k,a} = \begin{cases} +s \hat{x} &  k \in A \\ -s \hat{x} & k \in B \end{cases}\ , \quad
\hat\mu_{k,b} = \begin{cases} -s \hat{y} &  k \in A \\ +s \hat{y} & k \in B \end{cases}\ .
\ee
The parameter $s=\pm 1$ gives two distinct ground states. There is a $\pi$ phase shift for 
displacements along $x$ or $y$ through one primitive cell length.
This also means that a translation of the system through $\pm a\hat{x}$ or $\pm a\hat{y}$
will bring it to the opposite ground state.  There is no strong physical distinction of these 
two states, in terms of their physical properties.  But we can define order parameters 
to distinguish the two. 

An order parameter can be defined for each magnetic sublattice.  In the ground state,
the order is staggered, see Figure \ref{sqr-gs1}.  If the dipoles on the $B$ primitive cells are
inverted while the $A$ are left unchanged (i.e., apply an operation $R$ that is rotation 
through $\pi$ only on $B$), then all $A$ and $B$ primitive cells acquire the same configuration.  
Let the set $\hat\mu_{i}' = R_i\hat\mu_{i}$ be the configuration after this operation.  This 
operator $R$ has the definition,
\be
\label{R-def}
\hat\mu_{i}' = R_i \hat\mu_i = 
\begin{cases} +\hat\mu_i & \text{for $A$ cells} \\ 
              -\hat\mu_i & \text{for $B$ cells} \end{cases}
\ee 
Then we can define order parameters from sums on each magnetic sublattice:
\be
Z_a \equiv \frac{1}{N} \sum_{k} \hat\mu_{k,a}'\cdot \hat{x}, \quad
Z_b \equiv \frac{1}{N} \sum_{k} \hat\mu_{k,b}'\cdot \hat{y},
\ee  
where $N$ is the number of primitive cells.
The $s=+1$ ground state shown in Fig.\ \ref{sqr-gs1} has $Z_a=+1$ while $Z_b=-1$.  It is clear that these
are reversed in the other ground state.  Then the two can be combined into a single order
parameter for the whole system,
\be
Z = \frac{1}{2} \left( Z_a-Z_b \right).
\ee
This takes the values $\pm 1$ in the two ground states with $s=\pm 1$, respectively.  But 
by combining two parameters into one, some information has been forfeited. Further, this 
parameter does not contain information about the deviation of the dipoles sideways from 
$\hat{x}$ on the $a$-sites and from $\hat{y}$ on the $b$-sites.

One can further make vector order parameters, not the magnetization, but rather, using
the set of dipoles staggered by primitive cell, $\hat\mu_{i}'$.  These are defined on 
the magnetic sublattices:
\be
\label{w_ab}
\vec{w}_a = \frac{1}{N} \sum_k \hat\mu_{k,a}', \quad
\vec{w}_b = \frac{1}{N} \sum_k \hat\mu_{k,b}'.
\ee

Now consider that while not in the ground state, the Heisenberg-like dipoles are free to point
in any direction in three dimensions.  These vectors each have a component along an island
axis and a component transverse to the island axis.  The longitudinal parts give back the 
$Z_a$ and $Z_b$ parameters (not of unit magnitude):
\be
Z_a = \vec{w}_a \cdot \hat{x}, \quad Z_b=\vec{w}_b \cdot \hat{y}.
\ee
It is reasonable also to formulate a net vector order parameter for the system, from the sum,
\be
\vec{w} \equiv \frac{1}{\sqrt{2}} \left( \vec{w}_a + \vec{w}_b \right).
\ee
The factor $\tfrac{1}{\sqrt{2}}$ is for unit normalization, $|\vec{w}| = 1$, in a 
ground state.  The two ground states then correspond to $\vec{w}$ pointing along 
$-45^{\circ}$ [unit vector $\hat{v}=\tfrac{1}{\sqrt{2}}(\hat{x}-\hat{y})$]
and $+135^{\circ}$ [unit vector along $-\hat{v}$] from the $+\hat{x}$-axis.  

In a ground state, the $x$ and $y$ components of $\vec{w}$ are proportional to $Z_a$ and $Z_b$,
however, this is not true in an excited state, where generally we have
\bn
\label{wxwy}
w_x &=& \vec{w}\cdot\hat{x} = \frac{1}{\sqrt{2}}\left( Z_a + w_b^x \right), \nonumber \\
w_y &=& \vec{w}\cdot\hat{y} = \frac{1}{\sqrt{2}}\left( Z_b + w_a^y \right).
\en
The last parts, $w_b^x,\ w_a^y$, are the transverse elements, which vanish in a ground state. 
The overall order parameter $Z$ can be recovered from
\be
\frac{1}{\sqrt{2}}\left( w_x-w_y\right) = Z +\frac{1}{2} \left(w_b^x-w_a^y \right).
\ee
Thus the scalar order can be estimated from a slightly modified parameter
derived from the vector definition,
\be
\tilde{Z} = \vec{w}\cdot \hat{v}, \quad \hat{v}\equiv \tfrac{1}{\sqrt{2}}(\hat{x}-\hat{y}).
\ee
In a ground state, $\tilde{Z} = Z$, but otherwise they do not agree exactly, due to the
components contained in $\vec{w}$ that are transverse to the islands' long axes.

\begin{figure}
\includegraphics[width=\gsfigwidth]{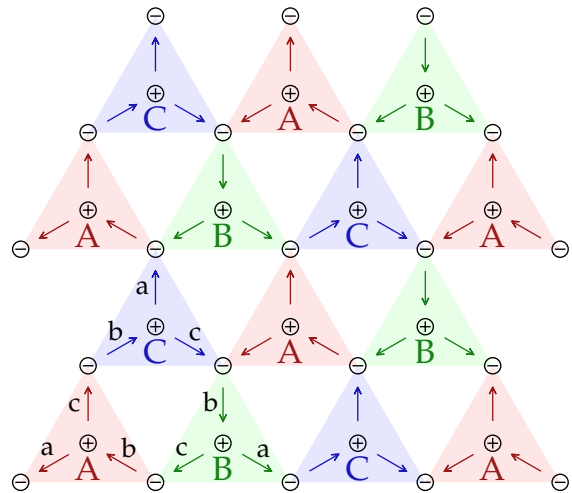}
\caption{\label{kag-gs1} Kagom\'e lattice spin ice in one of the ground states. A, B and C denote
the three sets of primitive cells. In each cell, the magnetic basis sites are a, b, c.  A rotation $R$
of the B cells through -120$^{\circ}$ and the C cells through +120$^{\circ}$ causes their a, b, c 
sites to align with those of the A cells. The circled plus and minus signs indicate singly-charged
magnetic monopoles.}
\end{figure}

\subsection{Order parameters for Kagom\'e ice}
A similar approach is followed to define order parameters for Kagom\'e ice.  The main 
difference is that the lattice consists of three types of primitive cells, labeled as 
A, B, C, that combine to make a 9-site magnetic unit cell.  Fig.\ \ref{kag-gs1} shows 
one of the ground states of the dipolar interactions.  The B and C cells can be considered 
copies of the A cells, but rotated by $120^{\circ}$ and $240^{\circ}$, respectively.  We 
consider the A cells as reference cells that get no rotation.  

In a ground state, if the B cells are rotated -120$^{\circ}$ around their centers, their 
a, b and c dipoles will fall into the corresponding positions and orientations of the A cells.
A +120$^{\circ}$ rotation of the C cells does the same for their dipoles.  This total operation 
applied to the whole lattice can be denoted $R$, such that the configuration becomes
\be
\hat\mu_i' = R_i \hat\mu_i = 
\begin{cases} +\hat\mu_i & \text{for $A$ cells} \\ 
               R(-\frac{2\pi}{3}) \cdot \hat\mu_i & \text{for $B$ cells} \\
               R(+\frac{2\pi}{3}) \cdot \hat\mu_i & \text{for $C$ cells} \end{cases} \ .
\ee
The operator $R(\pm\frac{2\pi}{3})$ does the $\pm 120^{\circ}$ rotation of an individual dipole.
Then each magnetic sublattice a,b,c, has a vector order parameter $\vec{w}_a,~\vec{w}_b,~\vec{w}_c$,
defined as in Eq.\ (\ref{w_ab}).  In the Ising model system, these $\vec{w}$ must align along the unit 
vectors that point outward from the A primitive cell centers,
\be
\hat{v}_a \equiv -\tfrac{\sqrt{3}}{2} \hat{x} -\tfrac{1}{2} \hat{y},\ \
\hat{v}_b \equiv +\tfrac{\sqrt{3}}{2} \hat{x} -\tfrac{1}{2} \hat{y},\ \
\hat{v}_c \equiv  \hat{y}\ .
\ee
The state in Fig.\ \ref{kag-gs1} satisfies  $\vec{w}_a = +\hat{v}_a,\ \vec{w}_b = -\hat{v}_b,\ 
\vec{w}_c = +\hat{v}_c$, which leads to a positive monopole in each primitve cell (and negative
monopoles at the cell junctions).  By permutations of the signs on $\hat{v}_a,~ \hat{v}_b,~\hat{v}_c$, 
it is easy to see that there are six ground states.   In analogy with square ice, the scalar
order parameters on the magnetic sublattices are
\be
Z_a = \vec{w}_a\cdot\hat{v}_a,\ \ 
Z_b = \vec{w}_b\cdot\hat{v}_b,\ \ 
Z_c = \vec{w}_c\cdot\hat{v}_c\ .
\ee
In a ground state, each of these is $\pm 1$, but with the sum $Z_a+Z_b+Z_c=\pm 1$, 
which is achieved in only six possible ways, for the six ground states.

On the other hand, a net vector order parameter is a sum,
\be
\vec{w} = \frac{1}{2}(\vec{w}_a+\vec{w}_b+\vec{w}_c)\ .
\ee
The normalization factor $1/2$ is useful so that $|\vec{w}|=1$ in a ground state.  Take the state
in Fig.\ \ref{kag-gs1}, which has $(Z_a,Z_b,Z_c)=(1,-1,1)$.  Because the magnetic unit vectors have
a net zero sum ($\hat{v}_a+\hat{v}_b+\hat{v}_c=0$), one gets for that state,
\be
\vec{w} = \frac{1}{2} ( \hat{v}_a - \hat{v}_b +\hat{v}_c ) = -\hat{v}_b\ .
\ee
In the ground states, this vector order parameter has the possible values, 
$\vec{w} = \pm\hat{v}_a,~ \pm\hat{v}_b,\ \pm\hat{v}_c$.  This is somewhat analogous
to a complex order parameter defined in other works (need ref).

An ordered state of the magnetic sublattices might be specified in an Ising system with
Ising variables $\sigma_a, \sigma_b, \sigma_c = \pm 1$, so that $\hat{\mu}_a = \sigma_a \hat{v}_a,~
\hat{\mu}_b = \sigma_b \hat{v}_b,~\hat{\mu}_c = \sigma_c \hat{v}_c,$.  This state
is specified by a vector,
\be
\Psi = \frac{1}{\sqrt{3}} ( \sigma_a, \sigma_b, \sigma_c ).
\ee
The ground state dipolar arrangements have state vectors,
\bn
\Psi_{\rm gs}^{1+} &\equiv& \frac{1}{\sqrt{3}} (1,-1,-1)\ , \nonumber \\
\Psi_{\rm gs}^{2+} &\equiv& \frac{1}{\sqrt{3}} (-1,1,-1)\ , \nonumber \\
\Psi_{\rm gs}^{3+} &\equiv& \frac{1}{\sqrt{3}} (-1,-1,1)\ .
\en
There are three others, $\Psi_{\rm gs}^{1-},~ \Psi_{\rm gs}^{2-},~ \Psi_{\rm gs}^{3-}$,
obtained from these by reversing the dipoles. Each Ising variable is $+1$ ($-1$) for outward (inward) from the
cell center.  These vectors clearly are overcomplete and not mutually orthogonal.  They have overlaps such as 
$\langle \Psi_{\rm gs}^{1+} | \Psi_{\rm gs}^{1+}\rangle=1$, 
$\langle \Psi_{\rm gs}^{1+} | \Psi_{\rm gs}^{2+} \rangle=-1/3$ 
and $\langle \Psi_{\rm gs}^{1+} | \Psi_{\rm gs}^{3+} \rangle=-1/3$.
The state shown in Fig.\ \ref{kag-gs1} is $\Psi_{\rm gs}^{2-}$.

In the general case we assume Heisenberg-like dipoles.  
On average, the sublattice magnetization vectors $\vec{w}_a,~\vec{w}_b,~\vec{w}_c$ can
point in any direction, then the Ising variables $(\sigma_a, \sigma_b, \sigma_c)$ of
an ordered magnetic state are replaced by projections $(Z_a, Z_b, Z_c)$, giving the
state variable,
\be
\Psi = \frac{1}{\sqrt{3}}\left( Z_a, Z_b, Z_c \right) \ .
\ee
A short calculation shows that this state is a superposition of the ground states,
\be
\Psi = Z_1 \Psi_{\rm gs}^{1+} + Z_2 \Psi_{\rm gs}^{2+} + Z_3 \Psi_{\rm gs}^{3+}
\ee
where the coefficients are combinations,
\be
\label{Z123}
Z_1= -\frac{Z_b+Z_c}{2},~ Z_2=-\frac{Z_a+Z_c}{2},~ Z_3=-\frac{Z_a+Z_b}{2} \ .
\ee
Therefore, specifiying the set $(Z_a,Z_b,Z_c)$ gives a sense of the order in each
sublattice, whereas, specifying the set $(Z_1,Z_2,Z_3)$ gives a sense of the condensation 
of the system into the possible ground states.  One should be aware, however, that
these $Z$ parameters do not contain the information about magnetization components 
that are transverse to the sublattice unit vectors $\hat{v}_a,~ \hat{v}_b,\ \hat{v}_c$.
A scalar product like $\vec{w}\cdot \hat{v}_a$ will approximate $Z_a$, however, the difference
is due to those transverse parts, as seen in the case of square ice, Eq.\ (\ref{wxwy}).
Note that the state in Fig.\ \ref{kag-gs1} has $(Z_1,Z_2,Z_3) = (0,-1,0)$, which shows it is
the pure $\Psi_{\rm gs}^{2-}$ ground state.

\subsection{Monopole charge densities}
A characteristic feature of two-dimensional spin ices is the presence of magnetic
monopole structures, that give a sense of the dipolar order or disorder in the system. 
At every primitive cell of the chosen lattice, there is a junction (or vertex) where a set of
dipoles can point outward or inward.  An imbalance in the net number of outward minus
inward dipoles corresponds to monopole charge.  In square ice,  a vertex has four dipoles,
while there are three dipoles at every vertex of Kagom\'e ice.  The coordination number for a 
vertex is $C=4$ in square ice and $C=3$ in Kagom\'e ice.

For each vertex at some position $\vec{r}_k$, one has a set of outward unit vectors,
$\hat{v}_{i_k},\ i_k=1,2,...C$  that point towards the nearest neighboring islands surrounding
that vertex.   One can define a discrete pole charge as has been used in square ice, counting
outward (inward) dipoles with $+1$ ($-1$) half-monopole contributions:
\be
\label{q-discrete}
q_k=\frac{1}{2} \sum_{i_k=1}^{C} [2H(\hat{\mu}_{i_k}\cdot \hat{v}_{i_k})-1]
\ee
where $H(x)$ is the Heaviside step function.  Then for square ice each vertex charge could have
the values $q=0,\pm 1, \pm 2$, depending on the local state of the four dipoles at that vertex. 
Thus, a vertex is either uncharged, singly-charged, or multiply-charged.  In
Kagom\'e ice, however, it is interesting to note that if one applies this {\em same} definition, the
poles are fractional.  The allowed values in Kagom\'e ice are $q=\pm \tfrac{1}{2}, \pm\tfrac{3}{2}$, 
due to the sum over only three inward or outward dipoles.  Even so, it will be convenient to refer to
the charges $q=\pm \tfrac{1}{2}$ as singly-charged poles and the $q=\pm \tfrac{3}{2}$ as 
multiply-charged poles.  In Kagom\'e ice the vertex charges are always nonzero, which makes
its thermodynamics rather distinct from square ice.

The discrete charge definition (\ref{q-discrete}) changes value suddenly as one dipole at a vertex
moves from pointing inward to outward or {\it vice-versa}.  Previously\cite{Wysin2013,Wysin2012} a continuously
varying charge-like definition was used, based only on scalar products,
\be
\label{q-contin}
q_k^{*} = \frac{1}{2} \sum_{i_k=1}^{C} \hat{\mu}_{i_k}\cdot \hat{v}_{i_k} .
\ee
This only sums the projections of the dipoles outward from vertex $k$.  A positive contribution 
in one vertex gives exactly the opposite negative contribution in a neighboring vertex.  Excluding the
boundary region of the system, the algebraic sum of all the $q$ is zero, either by this continuous version or the discrete version.  
For square ice, $q^{*}$ can take any value from $-2$ to $+2$, whereas, in Kagom\'e
ice the range is from $-\tfrac{3}{2}$ to $+\tfrac{3}{2}$.  Thus, the definition of continuous charge
$q^{*}$  parallels the definition of discrete charge $q$, although the extreme values like $q^{*}=\pm 2$ 
in square ice and  $q^{*}=\pm \tfrac{3}{2}$ in Kagom\'e ice fall only at the very end of the phase space, 
giving those points limited statistical weight.

The average magnetic charge density (combination of monopoles and multipoles) in the system is found by 
averaging the absolute valued charges over the $N_c$ vertices in the system,
\be
\rho = \langle \vert q \vert \rangle = \frac{1}{N_c} \sum_{k=1}^{N_c} \vert q_k \vert ,
\ee
with a similar definition for the continuous form, $\rho^{*}=\langle \vert q^{*} \vert \rangle$.  
The variation of these densities with temperature is a measure of entropic effects, but the
details in Kagome\'e ice are quite different than in square ice.

In the case of studying the discrete charges, we can also distinguish the singly-charged poles
from the multiply-charged poles, and count their averages separately in simulations.  Thus we
define partial densities denoted as $\rho_1$ and $\rho_2$ for square ice, that 
correspond respectively to the separate densities due to only singly and doubly charged poles.
Suppose the variables $n_1$ and $n_2$ represent the numbers of single and double charges at any 
vertex. Then the single and multiple densities are taken as $\rho_1=\langle n_1 \rangle$ and 
$\rho_2=2 \langle n_2 \rangle$.  The total charge density is then
\be
\rho = \rho_1 + \rho_2 = \langle n_1 + 2 n_2 \rangle.
\ee
Obviously, these relations account for that fact that doubly-charged poles contribute twice
as much charge as monopoles.  In Kagom\'e ice, we need instead to count single poles ($q=\pm \tfrac{1}{2}$)
and triple poles ($q=\pm \tfrac{3}{2}$), which have numbers $n_1$ and $n_3$ at sites.  Then with
single pole density $\rho_1=\tfrac{1}{2} \langle n_1 \rangle$ and triple pole density 
$\rho_3 = \tfrac{3}{2} \langle n_3 \rangle$, the total charge density can be obtained as
\be
\rho = \rho_1 + \rho_3 = \langle \tfrac{1}{2} n_1 + \tfrac{3}{2} n_3 \rangle.
\ee
Again, this accounts for the tripled charge on the multiply-charged poles in Kagom\'e ice.

It is also good to note another reason why the factor of $\frac{1}{2}$ is convenient for Kagom\'e 
ice as well as for square ice in the charge definitions (\ref{q-discrete}) and (\ref{q-contin}).  
In terms of the discrete charge definition, a unit of monopole
charge moves into one vertex from another when a single dipole of that vertex flips from inward 
to outward.  With the definition (\ref{q-discrete}) including the $\tfrac{1}{2}$, such a dipole
flip always contributes a change $\Delta q_k = \pm 1$ at a vertex in any lattice.  Thus, the flow of
monopolar charge within the system will be consistently counted in discrete units $\Delta q = \pm 1$
regardless of the type of lattice.

\subsubsection{Square ice charge densities.} \label{sqr-rho} For square ice, there are
no monopoles in either of the ground states ($q_k=q_k^{*}=0$ for all $k$), and $\rho=\rho^{*}=0$ for 
such an ordered state.  This can only be expected to be possible at very low temperature, 
assuming the system is able to access such a specific point of phase space.  Then, as 
the temperature is increased, small angular deviations of the dipoles away from the ground 
state directions will cause the individual $q_k^{*}$ charges to become nonzero, 
{\it before there are any significant changes in the discrete charges} $q_k$.  Thus, 
the continuous charge density $\rho^{*}$ initially will grow above zero more rapidly
with temperature than $\rho$, see Ref. \onlinecite{Wysin2013}.   

At high temperatures,
the system is of high entropy and in a configuration of maximum disorder.  Then, the
inidividual dipoles essentially become free to point in any direction on a unit sphere
(within the model of dipoles of fixed magnitude).  This allows calculation of the limiting
values of discrete and continuous charge density definitions.  In the discrete definition,
each of the four dipoles at a vertex is equally likely to point inward or outward. Of the
16 possible states, there are six with $|q|=0$, eight with $|q|=1$, and two with $|q|=2$.
The average gives the high-temperature limiting value,
\be
\rho =\langle \vert q \vert \rangle = \frac{1}{16}\left(6\times 0 + 8\times 1 + 2\times 2 \right) 
= \frac{3}{4}.
\ee
At any chosen vertex, the probability of finding a singly-charged monopole tends towards
$\langle n_1 \rangle = \tfrac{8}{16}=\tfrac{1}{2}$ (giving $\rho_1=\tfrac{1}{2}$), while the 
probability for a doubly-charged pole tends towards $\langle n_2 \rangle = \tfrac{2}{16}=\tfrac{1}{8}$
 (giving $\rho_2=\tfrac{1}{4}$). 

For the continuous definition, there is a corresponding averaging, but over outward projection
components $x_i=\hat\mu_i\cdot \hat{v}_i, i=1,2,3,4,$ that range uniformly from $-1$ to $+1$.
Then the high-temperature limiting value is found from an average
\bn
\rho^{*} &=& \langle \vert q^{*} \vert \rangle 
= \left\langle \frac{1}{2} \left\vert \sum_{i=1}^{4} x_i \right\vert \right\rangle \\
&=& \frac{\tfrac{1}{2} \int dx_1 \int dx_2 \int dx_3 \int dx_4 ~ \vert x_1+x_2+x_3+x_4 \vert}
  {\int dx_1 \int dx_2 \int dx_3 \int dx_4 ~  1 } = \frac{7}{15}.  \nonumber
\en
Thus, although the continuous form increases more rapidly at low temperature, its high-temperature
limit is actually less than that for the discrete definition.  Therefore the discrete and continuous 
charge definitions carry different kinds of information about the system and entropy effects.

\subsubsection{Kagom\'e ice charge densities.} For Kagom\'e ice, any of the ground states are 
ordered in such a  way that every vertex has a single charge: $q_k = q_k^{*}=\pm \tfrac{1}{2}$ 
for all $k$, and then $\rho =\rho^{*} = \tfrac{1}{2}$ at low enough temperature. Until the temperature
increases sufficiently to change away from the ground state configuration of monopoles, the discrete
charge density will remain at this value.  However, just as in square ice, the continuous
version of the charge density will already deviate from the value $\rho^{*}=\tfrac{1}{2}$ even
at very low temperatures, once there are angular deviations of the dipoles from their ground state
directions (radially outward/inward from the vertices).  

For high temperature the great entropy of the dipoles allows them to point equally in all
directions.  In the discrete charge definition inward and outward become equally probable.
For the three dipoles at a vertex, of their 8 possible states, there are six states with 
$|q|=\tfrac{1}{2}$ (single monopoles) and two states with $|q|=\tfrac{3}{2}$ (triple or multiple 
poles).  Then the high-temperature limiting density is the average,
\be
\rho =\langle \vert q \vert \rangle = \frac{1}{8}\left(6\times \tfrac{1}{2} + 2\times \tfrac{3}{2} \right) 
= \frac{3}{4}.
\ee
Note that with these fractional charges being used, the high-entropy charge density limit is
{\it the same as in square lattice ice}.  In this limit, every vertex has a probability of 
$\langle n_1 \rangle = \tfrac{6}{8}=\tfrac{3}{4}$ for singly-charged poles (giving $\rho_1=\tfrac{3}{8}$)
and a probability of $\langle n_3 \rangle = \tfrac{2}{8}=\tfrac{1}{4}$ for multiply-charged poles
(also giving $\rho_3=\tfrac{3}{8}$).  

For the continous charge definition at high-temperature, one now has an average over the
outward projections of only three dipoles, each ranging from $-1$ to $+1$.  Thus the limiting
value of continous charge density becomes
\bn
\rho^{*} &=& \langle \vert q^{*} \vert \rangle 
= \left\langle \frac{1}{2} \left\vert \sum_{i=1}^{3} x_i \right\vert \right\rangle \\
&=& \frac{\tfrac{1}{2} \int dx_1 \int dx_2 \int dx_3 ~ \vert x_1+x_2+x_3 \vert}
  {\int dx_1 \int dx_2 \int dx_3  ~  1 } = \frac{13}{32}.  \nonumber
\en
This is only slightly less than the limiting value in square ice.  This and the
other high-temperature limits are useful as checks on simulation results.

\subsection{Other correlations and measures}
For the purpose of analyzing geometric order in a spin ice, beyond the counting of monopole
charge, we can consider some short range correlations and various probability distributions.
What follows here is some discussion of other quantifiable local measures that give
certain information about the geometric states.

\subsubsection{Near neighbor correlations}
Generally, an ice system does not easily fall globally into one of its ground states, due
to the enormous geometrical frustration.  In a quench from higher temperature, one can expect
(at most) to have different regions close to the different ground states, with some connection
at the boundary between them, similar to domains connected by domain walls in magnets. 
This suggests looking at a more local order parameter, i.e., the nearest neighbor dipole
correlations, $\langle \hat\mu_i \cdot \hat\mu_j \rangle$, where $i$ and $j$ refer to any
two nearest neighbor dipoles.  

These correlations do not need to be within individual primitive cells.
The near neighbor correlations include intra-cell and inter-cell contributions. 
We can look at how the a,b,c magnetic sublattices are correlated with respect
to each other at the nearest neighbor distance.  

Consider first the correlations in Kagom\'e ice.
In one of the Kagom\'e ground states, if $i$ is on the a-sublattice with 
$\hat\mu_i = \hat\mu_a = Z_a \hat{v}_a$, then near neighbor site $j$ must be on either the 
b or c-sublattice, with a similar expression.  The correlation of this near neighbor pair 
(in a ground state) will be
\be
\langle \hat\mu_a \cdot \hat\mu_b \rangle = Z_a Z_b ~ \hat{v}_a\cdot \hat{v}_b = -\frac{1}{2} Z_a Z_b.
\ee
The factor of $-\tfrac{1}{2}$ is due to the 120$^{\circ}$ angle between $\hat{v}_a$ and $\hat{v}_b$.
Obviously there is a similar result for pairs on a and c sublattices and b and c sublattices.
These pair combinations then take values $\pm\tfrac{1}{2}$ in a ground state, depending on the
particular values of $Z_a, Z_b, Z_c$.  One can further see that some linear combinations of these
pair products take values of 0 or 1 in the ground states.  For instance, the sum
\be
\hat\mu_a \cdot \hat\mu_b + \hat\mu_a\cdot \hat\mu_c = -\frac{1}{2} Z_a (Z_b+Z_c)
\ee
takes the value 0 in the $\Psi_{\rm gs}^{2-}$ ground state shown in Fig. 2, with $(Z_a,Z_b,Z_c)=(1,-1,1)$,
and also in ground state $\Psi_{\rm gs}^{2+}$, but it is equal to +1 in all the other ground states.

The $Z$ parameters were defined as system averages, but generally we are concerned with local
variations. When the system is not in a ground state, we can still define quantities from the 
local correlations; the above discussion only suggests the limiting behavior.  In a short-hand 
notation, denote the sublattice-dependent correlations of nearest neighbors as
\bn
\label{Cab}
C_{ab} &=& \langle \hat\mu_a \cdot \hat\mu_b \rangle_{\rm nn}, \nonumber \\
C_{ac} &=& \langle \hat\mu_a \cdot \hat\mu_c \rangle_{\rm nn}, \nonumber \\
C_{bc} &=& \langle \hat\mu_b \cdot \hat\mu_c \rangle_{\rm nn}.
\en
The averages are over any near-neighbor pairs on the indicated sublattices.  
In any ground state of Kagom\'e ice, two of these are $+1$ and the 
third is $-1$, or, two are $-1$ and the other is $+1$.  From these basic correlations, we can 
then define combinations which give a sense of local proximity to the ground states,
\bn
\label{C123}
C_1 &=& \langle \hat\mu_a\cdot (\hat\mu_b+\hat\mu_c) \rangle = C_{ab}+C_{ac}, \nonumber \\
C_2 &=& \langle \hat\mu_b\cdot (\hat\mu_a+\hat\mu_c) \rangle = C_{ab}+C_{bc}, \nonumber \\
C_3 &=& \langle \hat\mu_c\cdot (\hat\mu_a+\hat\mu_b) \rangle = C_{ac}+C_{bc}.
\en
Unfortunately these last correlations do not distinguish between the pairs of ground states 
that are equivalent by a global inversion.  For instance, $C_1=0, C_2=+1, C_3=0$ in both of
the states, $\Psi_{\rm gs}^{2+}$ and $\Psi_{\rm gs}^{2-}$.  Clearly $C_2$ is defined
in such a way that it acquires the value of $|Z_2|$ in the limit of a ground state,
and similarly for $C_1$ and $C_3$.  It is important to note, however, that any of 
$C_1, C_2, C_3$ could, in principle, range from zero to 2 in magnitude.

For square ice, only $C_{ab}$ is defined, and it will be zero in a ground state, and
it will deviate away from zero at finite temperatures.  Instead, it may be better to consider a 
different correlation, defined so that it tends to unity in a ground state.  It is convenient
to define an operation $L$ that aligns all the dipoles of square ice in a ground state.
The operation that does this is a $+90^{\circ}$ rotation of the $b$-sites, after the 
$180^{\circ}$ rotations of $a$ and $b$ sites in the $B$ primitive cells (operation $R$ 
previously defined, Eq.\ \ref{R-def}).  This operation should be applied 
to the system before doing the correlations. When applied to the ground state in Fig. 1, 
all dipoles will align with the direction of the $a$-sites in the $A$ primitive cells.
The operation can be expressed using the cell-staggered $\hat{\mu}'$ dipoles 
[Eq.\ \ref{R-def}] by
\be
\label{L-def}
\hat{\mu}_i^{''} = L_i \hat{\mu}_i = 
\begin{cases}            \hat\mu_i' & \text{for $a$ sites} \\ 
               R(+\frac{\pi}{2}) \cdot \hat\mu_i' & \text{for $b$ sites} \end{cases}
\ee
The $+90^{\circ}$ rotation of an individual dipole is denoted $R(+\frac{\pi}{2})$.
In a short-hand notation, the near-neighbor correlation we have calculated for square 
ice is
\be
\label{Nab}
N_{ab} = \langle \hat{\mu}_a^{''} \cdot \hat{\mu}_b^{''} \rangle_{nn},
\ee
where the average is taken only over near neighbor pairs on opposite sublattices.
As stated, this average is unity in one of the ground states, and less than unity
as the systems moves away from a ground state.

\subsubsection{Probability distributions for ice order}
The parameters $Z_a,Z_b,Z_c$ were defined as averages  over the whole system, of components
of the dipoles along the sublattice unit vectors $\hat{v}_a,\hat{v}_b,\hat{v}_c$.
Of course, these averages can be considered as derived from distributions of some
local variables, $z_{a},z_{b},z_{c}$, with local definitions,
\be
\label{localza}
z_{a}=\hat\mu_{k,a}'\cdot \hat{v}_a,\quad
z_{b}=\hat\mu_{k,b}'\cdot \hat{v}_b,\quad
z_{c}=\hat\mu_{k,c}'\cdot \hat{v}_c,
\ee
all obtained from the dipoles of a cell $k$.  These components range from $-1$ to $+1$.
Due to thermal fluctuations the values will vary from cell to cell, and also over time.
Therefore, in simulations of the dynamics they can be periodically grouped into bins
from which their probability distributions can be estimated.  For the simulations
presented in this work we partitioned the range $-1 \le z \le +1$ into 201 bins.  In 
C-language coding, an array $p[201]$ for one of these distributions can be updated by 
an algorithm such as
\be
p[ 100*(z+1) ]\ +=\ 1;
\ee
and afterwards normalized to unit total area.  By running a
simulation and accumulating data, this allowed for the calculation of probabilities
$p_a(z_a), p_b(z_b), p_c(z_c)$ corresponding to each sublattice's order.  These can be
considered the raw probability distributions for the local order.  In a ground state,
these distributions become highly skewed towards the appropriate extreme values 
$z_a\rightarrow \pm 1$, $z_c\rightarrow \pm 1$, $z_c\rightarrow \pm 1$.

In square lattice ice, a derived probability distribution can be found for the net local
order parameter
\be
\label{localz}
z= \frac{1}{2} (z_a-z_b).
\ee
The probability $p(z)$ is obtained indirectly from $p_a(z_a)$ and $p_b(z_b)$ by a relation,
\be
p(z) = \int dz_a ~ p_a(z_a) \int dz_b ~ p_b(z_b) ~ \delta[z-\tfrac{1}{2}(z_a+z_b)].
\ee
This was implemented by coding an equivalent C-language algorithm for the probability arrays.

In Kagom\'e lattice ice, a similar approach can be used to find derived probability
distributions that measure proximity to the ground states.  For example, generalizing
the coefficients in (\ref{Z123}) to local definitions,
\be
\label{z123}
z_1 = -\frac{z_b+z_c}{2},\quad 
z_2 = -\frac{z_a+z_c}{2},\quad 
z_3 = -\frac{z_a+z_b}{2},
\ee
then the probability distribution of $z_1$ is found from
\be
\label{localz1}
p_1(z_1) = \int dz_b ~ p_b(z_b) \int dz_c ~ p_c(z_c) ~ \delta[z_1+\tfrac{1}{2}(z_b+z_c)].
\ee
There are similar ways to get $p_2(z_2)$ and $p_3(z_3)$.  These three distributions
will now give a graphic view of the extent of condensation of the system to regions
of phase space locally near the respective ground states.

\begin{figure}
\includegraphics[width=\figwidth,angle=-90.]{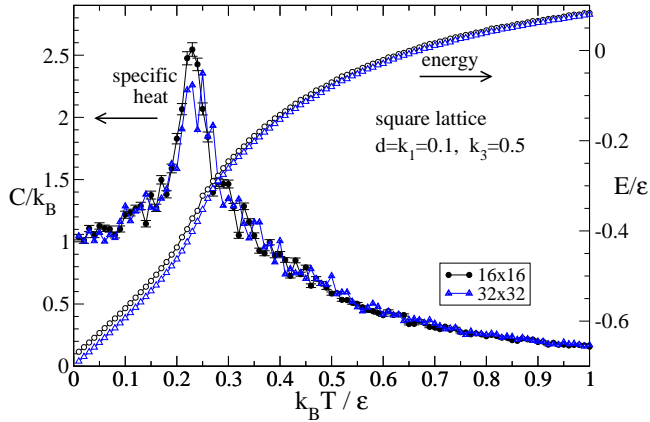}
\caption{\label{sqr-eckT} The specific heat and internal energy per site for square lattice 
spin ice with artifical model parameters. Clear evidence of a phase transition appears as the
peak in specific heat near $k_B T/\varepsilon = 0.22$ .}
\end{figure}

\begin{figure}
\includegraphics[width=\figwidth,angle=-90.]{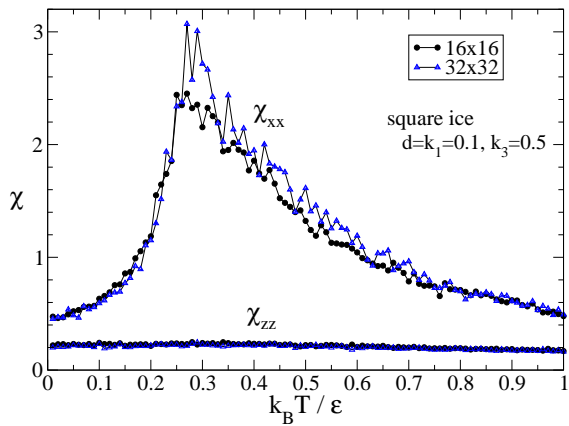}
\caption{\label{sqr-chikT} For square lattice spin ice, the behavior of the
per-site magnetic susceptibility components with temperature.}
\end{figure}

\begin{figure}
\includegraphics[width=\figwidth,angle=-90.]{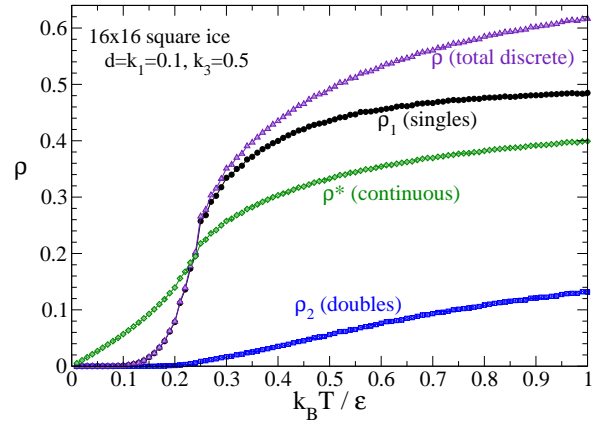}
\includegraphics[width=\figwidth,angle=-90.]{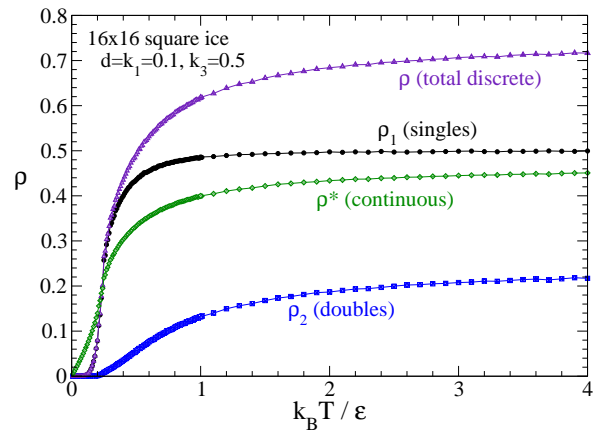}
\caption{\label{sqr-qnkT} The temperature dependence of monopole charge densities for 
the $16 \times 16$ square lattice spin ice articifical model. The discrete total density $\rho$
is the sum of single and double pole charge densities: $\rho=\rho_1+\rho_2$. The continuous
definition for $\rho*$ is more a measure of the deviations of the dipoles away from a ground
state, where $\rho=\rho*=0$. Similar results were obtained for a $32 \times 32$ system.
Part (a) shows the low-temperature dependence and the phase transition near $k_B T/\varepsilon \approx 0.22$ .  
Part (b) exhibits the expected high-temperature asymptotic values $\rho_1=\tfrac{1}{2},\ \rho_2=\tfrac{1}{4}$
explained in the text.}
\end{figure}

\section{Simulations of thermal ordering in spin ices}
In this work we use study the thermal equilibrium dynamics of an array of dipoles making
up the spin ice.  To give realistic dynamics, the torques that tend to align the dipoles
along island axes are modeled by including appropriate uniaxial magnetic anisotropies in
the spin model.  For this two-dimensional (2D) array of very thin nanoislands, uniaxial 
energy parameter $K_1$ determines the strength of alignment along the island axes.
Another energy parameter $K_3$ is used to make the $z$-axis perpendicular to the plane
of the islands a hard axis.  The dipoles themselves are of three components, and although
free to point in any direction, restricted by these local torques that are used to
model the geometrical anisotropies of the islands.  Including also a long-range dipolar
interaction, the Hamiltonian of the ice system is
\bn
{\cal H} &=& -\frac{\mu_0}{4\pi} \frac{\mu^2}{a^3}
\sum_{i>j} \frac{ \left[ 3(\hat{\mu}_i\cdot \hat{r}_{ij})(\hat{\mu}_j\cdot\hat{r}_{ij})
                                -\hat{\mu}_i\cdot \hat{\mu}_j \right]}
{\left( {r}_{ij} / a\right)^3} \\
&+&  \sum_{i} \left\{ K_1[1-(\hat\mu_{i}\cdot\hat{u}_i)^2] + K_3 (\hat\mu_{i}\cdot \hat{z})^2
-\vec\mu_i \cdot \vec{B}_{\rm ext} \right\}
\nonumber
\en
The unit vectors $\hat{r}_{ij}$ point from site $j$ to site $i$, and $\mu_0$ is the magnetic permeability 
of free space.  The dipoles themselves have assumed fixed  magnitude $\mu$ while being able to rotate to 
different directions.  The relative strength of dipolar interactions (first term) depends on the
lattice constant for the ice array, $a$, equal to the distance between any nearest neighbor pair 
of dipoles.  That dipolar energy parameter is
\be
D = \frac{\mu_0}{4\pi} \frac{\mu^2}{a^3}.
\ee
The unit vectors $\hat{u}_i$ are the fixed island anisotropy axes.  Each is equivalent
to one of the choices of $\hat{v}_a, \hat{v}_b$ or $\hat{v}_c$, depending on the magnetic sublattice
of that site.  The very last term includes an applied external magnetic field.

The relative sizes of the different physical effects can be indicated by their dimensionless
energy constants, scaled in units of the basic energy scale, 
\be
\varepsilon \equiv \mu_0 \mu M_s,
\ee
where $M_s$ is the saturation magnetization of the island material.  Then the dimensionless
energy parameters that define the problem are
\bn
\label{dk1k3}
d &=& \frac{D}{\varepsilon} = \frac{\mu}{4\pi a^3 M_s},
\\
k_1 &=& \frac{K_1}{\varepsilon}, \quad
k_3 = \frac{K_3}{\varepsilon}.
\en
From this, it is apparent that dipolar effects are directly determined by the volume fraction
of the system occupied by magnetic material, since $\mu = M_s V_{\rm island}$ is proportional 
to each island's volume, $V_{\rm island}$.  When the parameter $d$ is very small, dipolar effects
can be ignored.  However, the presence of adequately strong $k_1$ will lead to freezing of
the dynamics; this would be the case of Ising-like anisotropy that limits dipolar rotation.

\begin{figure}
\includegraphics[width=\figwidth,angle=-90.]{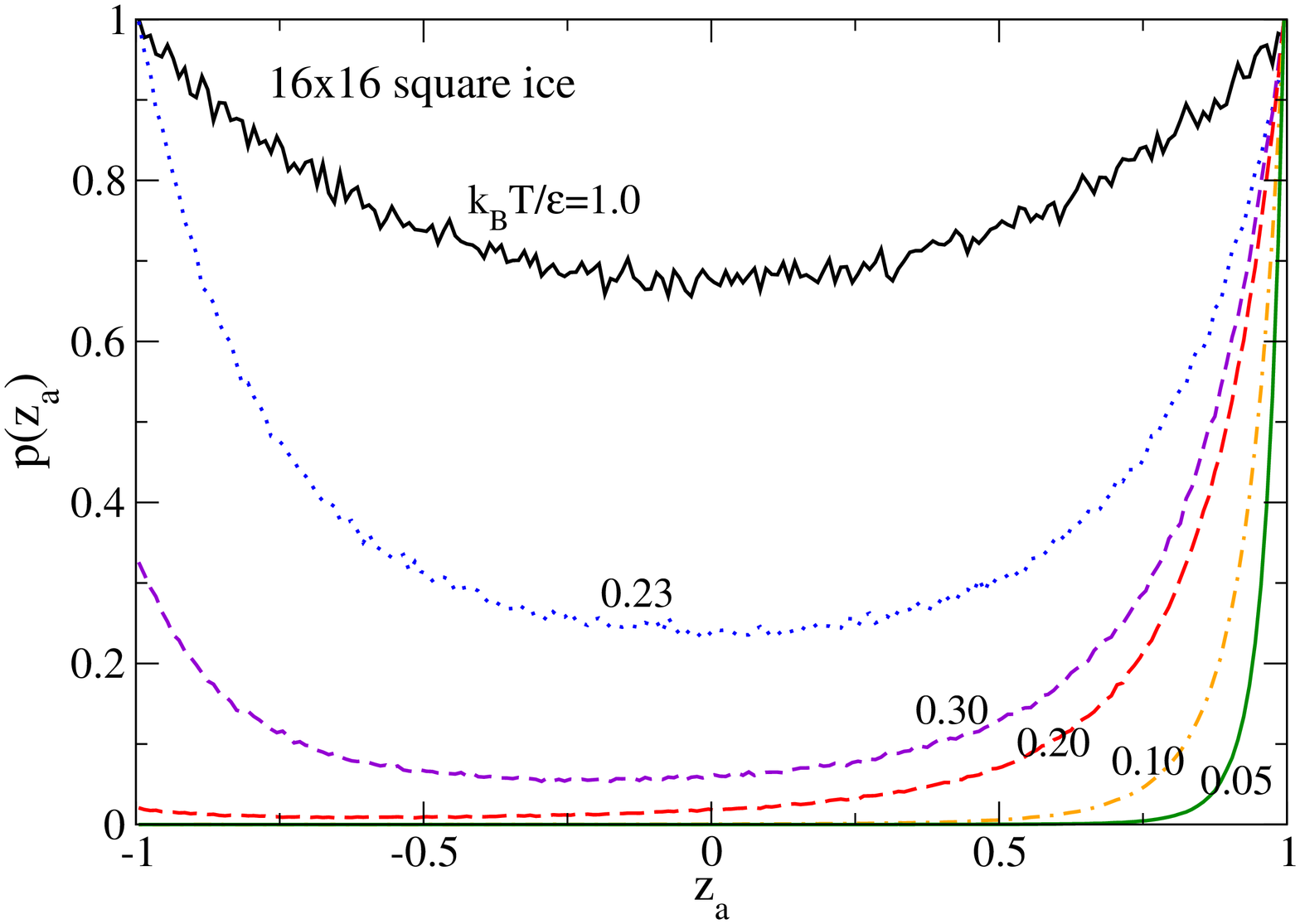}
\includegraphics[width=\figwidth,angle=-90.]{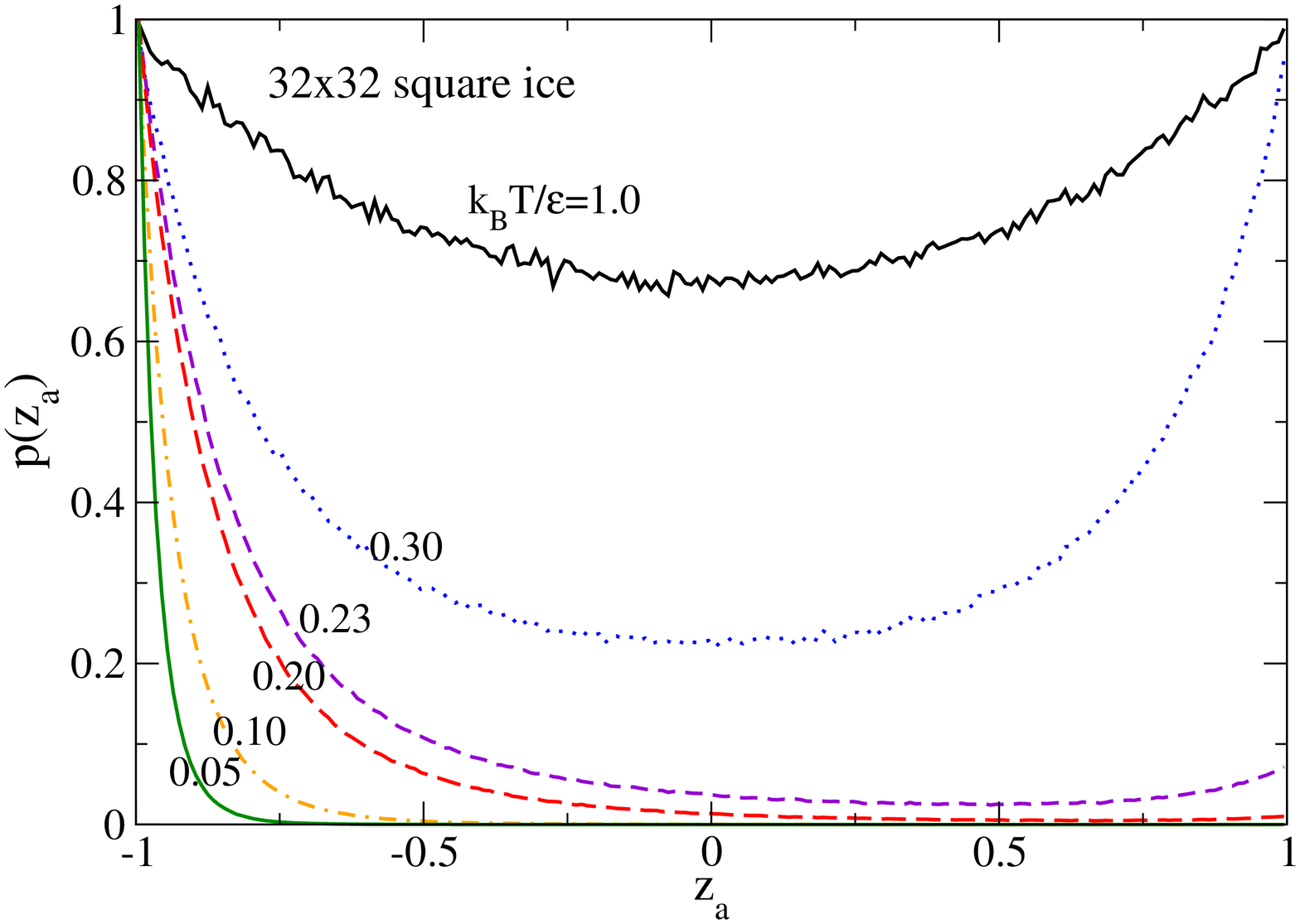}
\caption{\label{sqr-pza} The raw probability distribution of a-sublattice order $z_a$, 
Eq.\ (\ref{localza}), for square spin ice in (a) a $16 \times 16$ system and (b) 
a $32 \times 32$ system.  Labels by curves indicate the dimensionless temperature.
The distribution on the sublattice clearly displays randomness at high temperature
but strong tendency of alignment as the temperature is lowered. Similar behavior
takes place for $z_b$.}
\end{figure}

\begin{figure}
\includegraphics[width=\figwidth,angle=-90.]{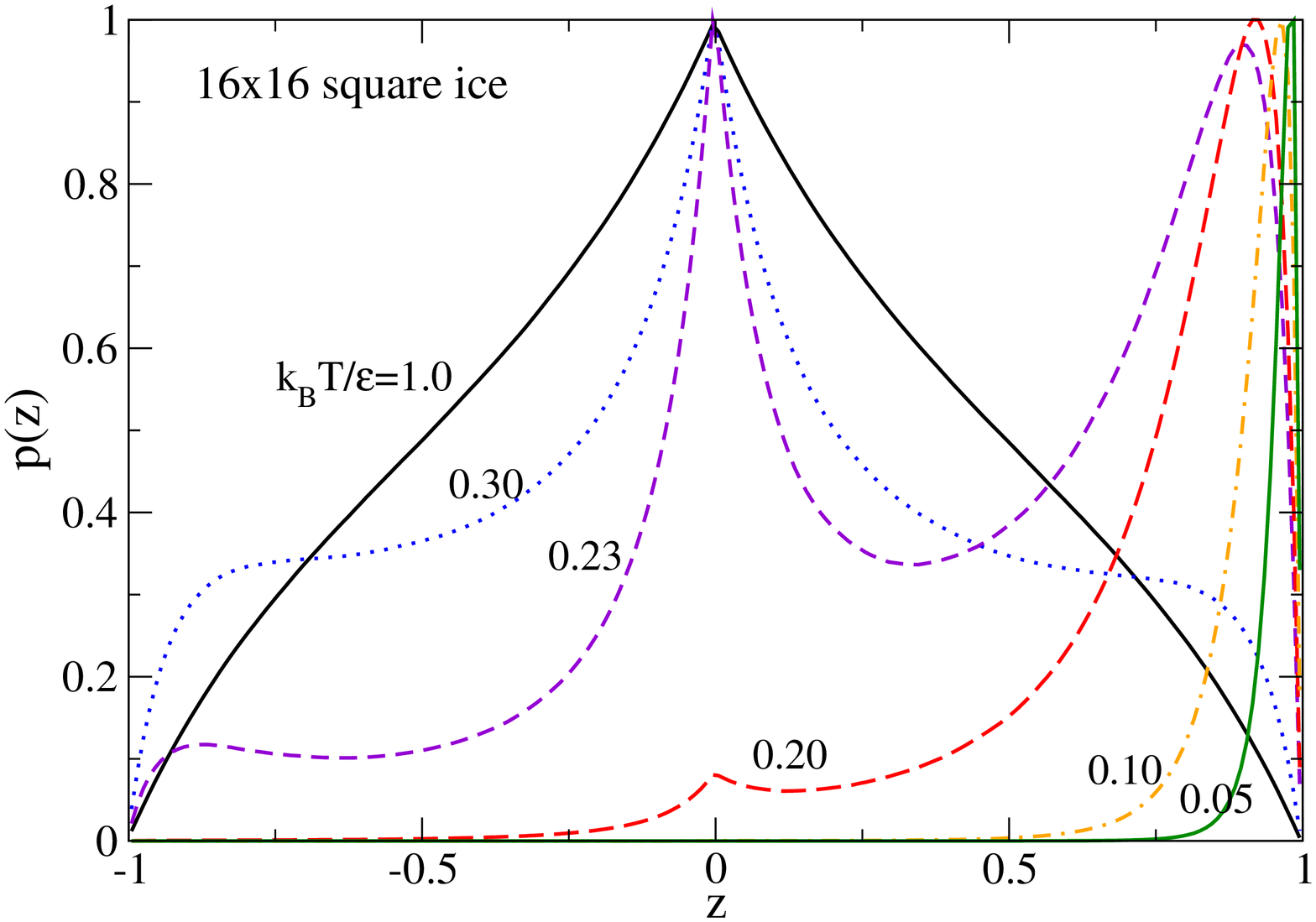}
\includegraphics[width=\figwidth,angle=-90.]{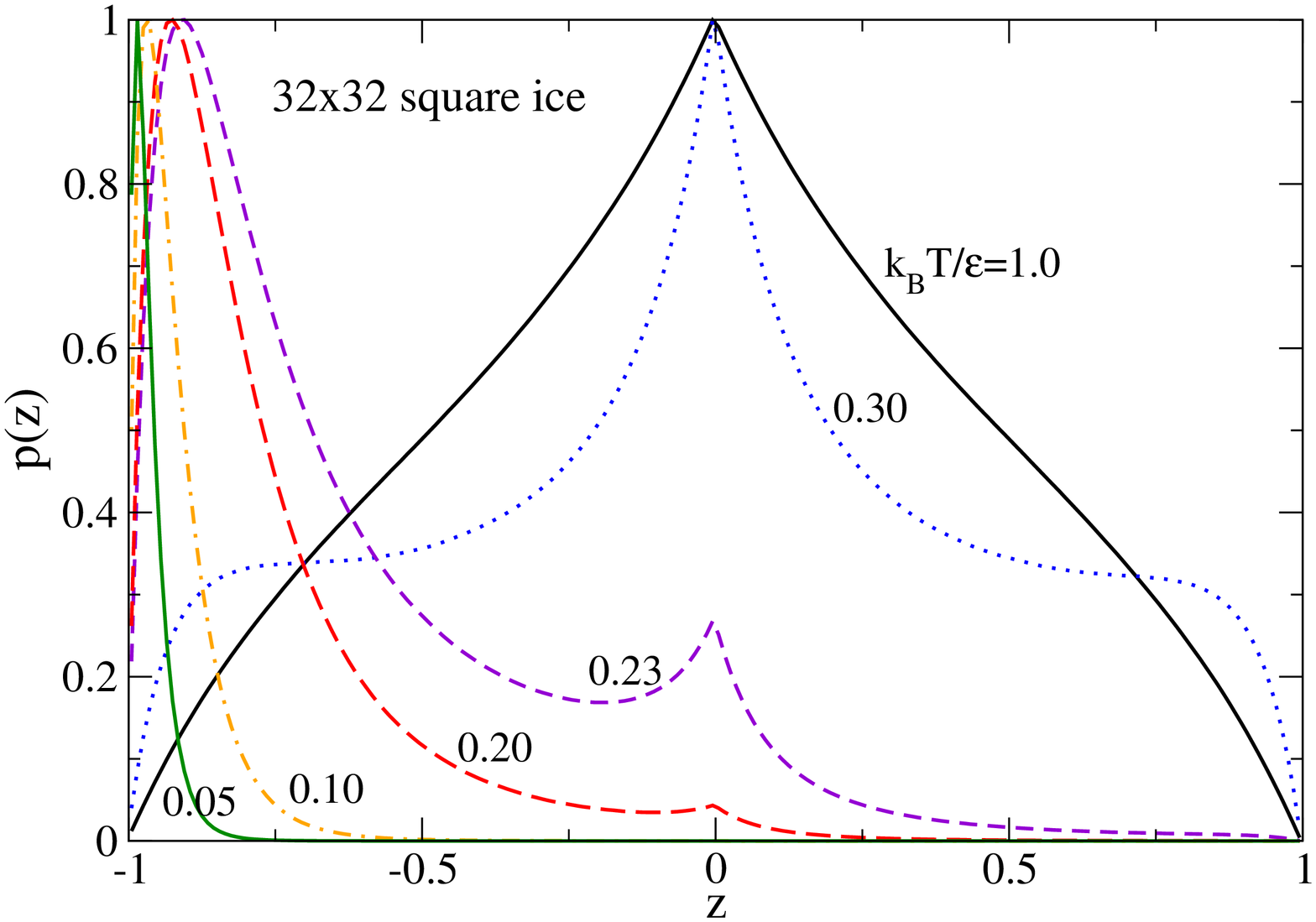}
\caption{\label{sqr-pz} Examples of the derived probability distribution of the local order
parameter $z$, Eq.\ (\ref{localz}), for square lattice spin ice, in (a) a $16 \times 16$ 
system and (b) a $32 \times 32$ system.  Different curves are labeled by the dimensionless 
temperature.  The distribution changes from wide at high $T$ (with $\langle z \rangle \approx 0$) 
to very skewed at low $T$ (with $\langle z \rangle \rightarrow \pm 1$).}
\end{figure}

Starting from an appropriate initial state, the dynamics of this system was solved according
to a Langevin equation derived from the Landau-Lifshitz-Gilbert equation for the dynamics.
The undamped zero-temperature dynamic equation can be written in dimensionless quantities as
\be
\frac{d\hat\mu_i}{d\tau} = \hat\mu_i \times \vec{h}_i, \quad
\tau = \gamma \mu_0 M_s\, t.
\ee
where $\gamma$ is the electron gyromagnetic ratio and $\vec{h}_i = \vec{B}_i / \mu_0 M_s$ 
is the dimensionless effective magnetic field that acts on the dipole at site $i$.
The local induction $\vec{B}_i$ is derived from the given Hamiltonian.  The Langevin 
equation includes damping and stochastic torques into
these equation, whose strength is related to the desired temperature.  
The Langevin equation was solved by a 2nd order Heun algorithm. 
The details of this procedure have been presented in Ref.\\onlinecite{Wysin+13}. 

As far as the choice of parameters, we consider a model system with parameters selected
so that the physical effects of frustration do not dominate excessively over thermal effects.
The difficulty is that the energy scale $\varepsilon$ tends to be much larger than thermal
energy even at typical room temperature, unless the islands are extremely small.  Instead,
we suppose an artificial choice of parameters so that the system is not too strongly Ising-like.  
This will ensure a thermalized dynamics for room temperature and somewhat below room temperature. 
We do this to avoid a situation where the uniaxial anisotropy is so strong that it prevents 
motion of the dipoles.  The intention is to see the realistic spin dynamics of the dipoles.  
Although it may be difficult to prepare an ice array with this choice, we choose model 
parameters $D=K_1=0.1 \varepsilon$,  $K_3=0.5 \varepsilon$, similar to Model C in 
Ref.\ \onlinecite{Wysin+13}, but with stronger easy-plane interaction.  In order to 
discuss results, the temperature is scaled with the same energy unit $\varepsilon$,
so that we quote dimensionless temperatures,
\be
{\cal T} = k_B T /\varepsilon,
\ee
where $k_B$ is Boltzman's constant.

\section{Results for Square Lattice Ice}
For square ice the dynamics was tested on systems of sizes $16\times 16$ and $32\times 32$,
with open boundaries.  These simulations represent the dynamics in  a small piece of a finite 
spin ice.  For square ice the calculation of the dipolar interactions is accelerated by use of
the FFT approach.   The damping parameter is set to $\alpha = 0.1$ for the Langevin dynamics. 
To run the dynamics, the basic time step for the second-order Heun algorithm is set to 
$\Delta \tau = 0.001$, but data for analyzing dynamics is sampled every 1000  Heun
steps, i.e., $\Delta \tau_{\rm samp} = 1000 \Delta \tau$.   The first 400 (200) samples are thrown 
out to equilibrate the system for the chosen temperature, and then averages are calculated
from the subsequent 4000 (2000) time samples for the $16 \times 16$ ($32\times 32$) system.  
This final time corresponds to many times the natural period of the dominant dipolar
oscillations.  Less samples were used for the larger system as a practical matter,
due to the considerably longer computation time. In a typical run over a set of temperatures, 
the highest temperature is calculated first, and the final state obtained there becomes the 
initial state for the next lower temperature.  Some typical final states are shown in 
Appendix A, Fig.\ \ref{sqr-configs}.

Results for the temperature dependence of specific heat and energy per dipole are given in 
Fig.\ \ref{sqr-eckT}.  As is well known in this type of ice model, a strong peak in the specific heat
near $k_B T \approx 0.22 \varepsilon$, indicative of a phase transition.  It is expected that
the low-temperature phase is well-ordered and bears at least a resemblance to one of the
ground states, when viewed at a local level.  

Although not as dramatic as the peak in specific heat, there is a broad peak in the
in-plane magnetic susceptibility versus temperature, see Figure \ref{sqr-chikT}.  The 
position depends somewhat on system size, but appears to be near $k_B T / \varepsilon \approx 0.27 $.
As can be expected, the in-plane magnetic fluctuations and susceptibility are much greater than
those out of the easy plane.  However, both $\chi_{xx}$ and $\chi_{zz}$ tend towards nonzero
values in the limit of zero temperature.

A further indication of the microscopic state
can be viewed in the temperature dependence of the various monopole charge densities, see
Fig. \ref{sqr-qnkT}.  For high temperatures, there is strong disorder, and the charge densities
obtained tend to the asymptotic values in Sec.\ \ref{sqr-rho}, specifically, $\rho_1\rightarrow 0.5$
for single poles and $\rho_2\rightarrow 0.25$ for double poles.  One can see, however, that
the approach to this high-disorder limit is very slow with increasing temperature. At the
low temperature range, in contrast, the density of single poles surges upward near 
$k_B T \approx 0.2 \varepsilon$, and double poles start to appear at a slightly higher temperature.
The highest slope $d\rho_1/dT$ takes place close to $k_B T \approx 0.22 \varepsilon$, the
same as the location of the peak in specific heat.  This shows that generation of pole density 
is responsible for the strong thermodynamic changes in the phase transition.

The degree of geometric ordering of the dipoles can also be seen in the probability
distributions for local order parameters $z_a$, $z_b$, and the derived $z$, see Eq.\
(\ref{localz}).  At higher temperature, Fig.\ \ref{sqr-pza} shows how there is a great
deal of randomness in the distribution of $z_a$, as can be expected.  The data
there and corresponding data for $p(z_b)$ (not shown) lead to the results in Fig.\
\ref{sqr-pz} for the net local order parameter distribution, $p(z)$.  The distribution $p(z_a)$
is broad with a nearly symmetric appearance plus noticable fluctuations at high temperature.
The somewhat parabolic form centered on $z_a=0$ can be explained by a distribution approximated
as $p(z_a) \propto \exp\{K_1 z_a^2/k_B T\}$, considering that the uniaxial energy would be the dominant 
interaction for the high-temperature disordered phase.  This gives a more uniform distribution
as the temperature increases, but yet reasonably explains the curvature of the function.
For lower temperatures the distribution transforms to a skewed form that concentrates towards one of
the end points, $z_a \rightarrow \pm 1$.  The same behavior holds for the distribution on the
other sublattice, $p(z_b)$.  The resulting derived distribution for $p(z)$ similarly becomes
skewed towards one of the end points for low temperature, see Fig.\ \ref{sqr-pz}, as the
system moves close to one of the particular ground states. 

\begin{figure}
\includegraphics[width=\figwidth,angle=-90.]{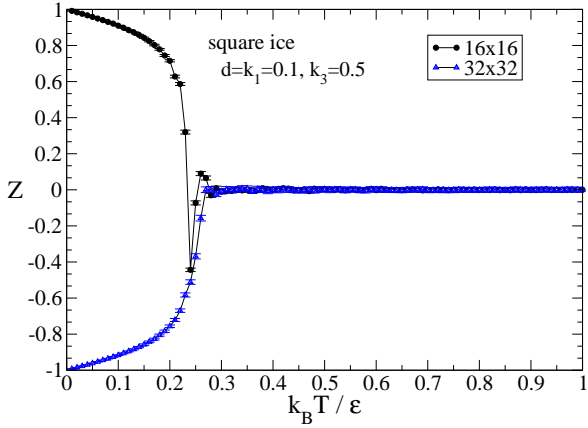}
\caption{\label{sqr-ZkT} Temperature dependence of the system order parameter $Z$ for square lattice spin 
ice, for two different system sizes.  The emergence into $Z \rightarrow \pm 1$ at low temperature
is a random choice between the two ground states.}
\end{figure}

\begin{figure}
\includegraphics[width=\figwidth,angle=-90.]{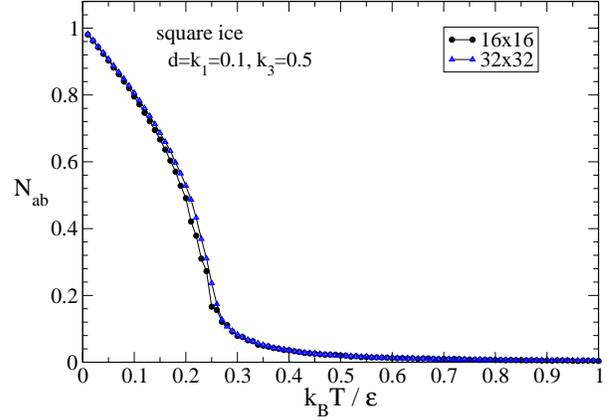}
\caption{\label{sqr-CabkT} Inter-sublattice nearest neighbor correlations [Eq.\ (\ref{Nab})]
for the square lattice spin ice model.}
\end{figure}

For the overall system order, the averaged parameter $Z$ displays the behavior with temperature
shown in Fig.\ \ref{sqr-ZkT}.  $Z$ displays a drastic drop to zero for temperatures $k_B T > 0.25 \varepsilon$.
At low temperature, it tends linearly towards $\pm 1$;  these two limiting values clearly have equal
likelihood and reflect the projection into one of the ground states.  

The nearest neighbor correlations in square ice, as represented by $N_{ab}(T)$, take the behavior 
shown in Fig.\ \ref{sqr-CabkT}.  In a certain sense the shape of the curve mimics the behavior of 
$Z(T)$.  There is a linear approach towards $N_{ab} = 1$ in the low temperature region, showing 
the approach into a ground state.  Note that this correlation function takes the value $N_{ab}=1$ 
in both of the ground states.  The function begins to deviate from linear behavior already by 
$k_B T \approx 0.2 \varepsilon$.  $N_{ab}$ does not go dramatically to zero, but rather, exhibits 
a very long decay towards zero starting at temperatures around $k_B T \approx 0.25 \varepsilon$.  
This indicates the generally disordered states far from the ground state configuration that
dominate the thermodynamics at high temperature.

\begin{figure}
\includegraphics[width=\figwidth,angle=-90.]{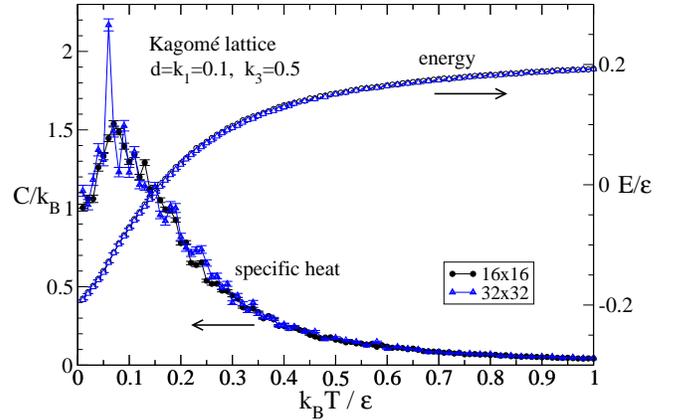}
\caption{\label{kag-eckT} The specific heat and internal energy per site for the 
Kagom\'e lattice spin ice model with artificial parameters. The transition from
disordered to ordered occurs around $k_B T / \varepsilon \approx 0.07$ .  Data
were generated by taking the system from high to low temperature.}
\end{figure}

\begin{figure}
\includegraphics[width=\figwidth,angle=-90.]{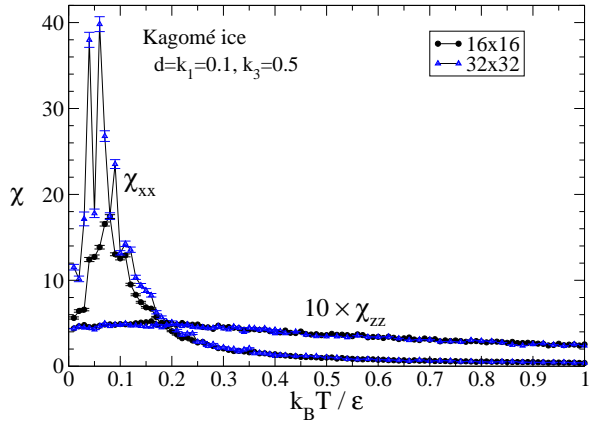}
\caption{\label{kag-chikT} The magnetic susceptibility components for the Kagom\'e lattice 
spin ice model, as obtained from fluctuations of the magnetization.  With the in-plane component 
dominating, the out-of-plane susceptibility has been scaled up by a factor of $10$.}
\end{figure}

\section{Results for Kagom\'e Lattice Ice}
For spin ice on a Kagom\'e lattice, the dynamics was also tested on systems of sizes 
$16 \times 16$ and $32 \times 32$ with open boundaries.  We used the same energy
parameters, $d=k_1=0.1$ and $k_3=0.5$,  that is, similar to Model C in Ref.\ \onlinecite{Wysin+13}.
The parameters for running the dyanmics (time step, sampling rate, total samples, etc.) were set 
to the same values as those for the square ice simulations.  However,  due to the more complex 
lattice structure, the dipolar interactions were taken into account by a numerical sum over
the whole system, rather than the accelerated FFT approach.  This means the execution is slower.  
As for square ice, most simulations were carried out by starting at the highest temperature and
using the final state of one temperature as the initial state of the next lower temperature.
Some typical final states are shown in Appendix A, Fig.\ \ref{kag-configs}.

Results for the specific heat and internal energy per dipole are displayed in Fig.\ \ref{kag-eckT}.
For Kagom\'e ice the transition to a more ordered state occurs near the temperature 
$k_B T/\varepsilon \approx 0.07$, about one third of the transition temperature
in square ice with the same interaction parameters.  There can be two reasons for this.
First, a dipole in the Kagom\'e system has only three nearest neighbors, compared to four
for square ice.  This reduces the effective strength of dipolar interactions and their ability
to bring the system into an ordered state.  Second,  the Kagom\'e system has a much
stronger geometric frustration that prevents it from condensing easily into one of its
ground states.  Of course, with six ground states to choose from, and these related to 
each other by fairly simple rotational symmetries, it is nearly impossible to move into
one of them over the whole extent of the system.  Therefore, the low-temperature phase
obtained in these simulations of cooling the system from higher temperatures leads to 
strong frozen-in disorder.  This is confirmed by the other measures we have calculated, 
as discussed below.

The temperature dependence of magnetic susceptibility components is shown in Fig.\ \ref{kag-chikT}.
The in-plane component $\chi_{xx}$ shows a strong peak at approximately the same temperature where
the peak in specific heat occurs.  There may be a finite size effect, as the peak is stronger 
in the larger system.  The out-of-plane component $\chi_{zz}$ shows only a very broad plateau,
while being considerably weaker than $\chi_{xx}$.  This difference in magnitudes is due to
the large value of planar anisotropy ($k_3=0.5$) relative to the uniaxial anisotropy ($k_1=0.1$),
that limits the size of $z$-components of the dipoles.

\begin{figure}
\includegraphics[width=\figwidth,angle=-90.]{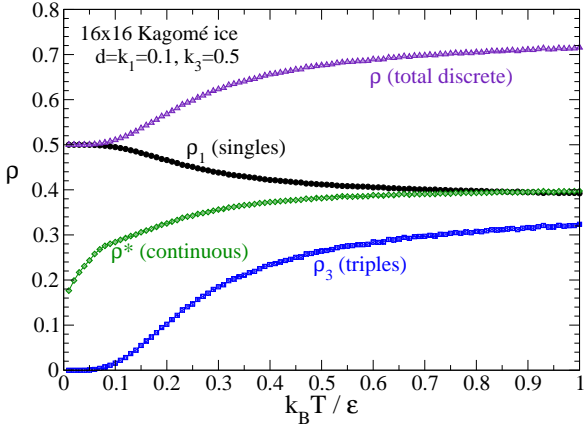}
\includegraphics[width=\figwidth,angle=-90.]{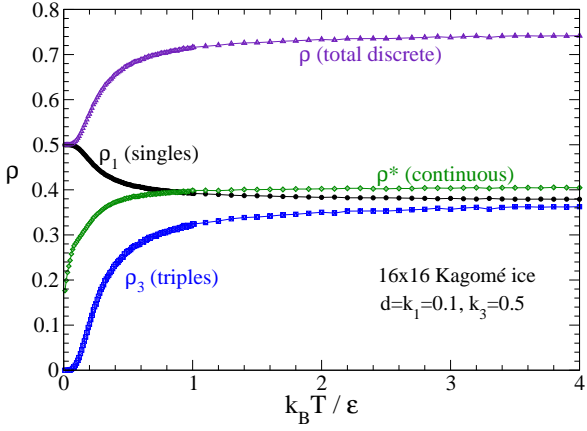}
\caption{\label{kag-qnkT} Monopole charge densities in $16\times 16$ Kagom\'e lattice spin ice as
functions of temperature. Similar results were obtained for a $32 \times 32$ system.
Total discrete pole density is the sum of single poles ($q=\pm \tfrac{1}{2}$) and triple
poles ($q=\pm \tfrac{3}{2}$).  Part (a) shows the low-temperature dependence and phase transition 
near $k_B T/\varepsilon \approx 0.07$ .  Part (b) shows the behavior going towards the high-temperature
asymptotic values $\rho_1=\rho_3=\tfrac{3}{8}$ discussed in the text.}
\end{figure}

\begin{figure}
\includegraphics[width=\figwidth,angle=-90.]{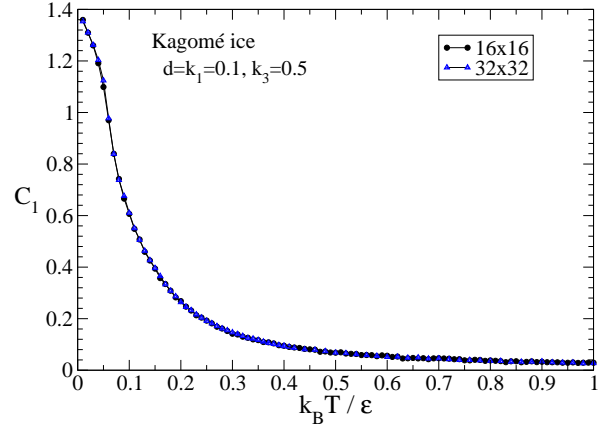}
\caption{\label{kag-C1kT} Averaged near neighbor correlations $C_1(T)$ for Kagom\'e lattice spin 
ice, as defined in Eq.\ (\ref{C123}).  The results for the other correlations $C_2(T)$ and $C_3(T)$
are essentially the same. If the system condensed into either the $\Psi_{\rm gs}^{1+}$ or
$\Psi_{\rm gs}^{1-}$ ground states, one would obtain $C_1=1, C_2=C_3=0$.}
\end{figure}

An indication of the behavior of the nearest neighbor correlations in Kagom\'e ice is
shown in Fig.\ \ref{kag-C1kT}, where $C_1(T)$ as defined in Eq.\ (\ref{C123}) is plotted.
The correlation $C_1(T)$ gives a measure of local projection onto the $\Psi_{\rm gs}^{1 \pm}$
ground states, but only if taken in conjunction with $C_2(T)$ and $C_3(T)$.  That is, if all
primitive cells of the system go to the state $\Psi_{\rm gs}^{1+}$ or $\Psi_{\rm gs}^{1-}$, 
then the values $C_1=1, C_2=0, C_3=0$ result.  In this simulation, however, all three measures,
$C_1(T)$, $C_2(T)$, and $C_3(T)$ take the temperature dependence seen in Fig.\ \ref{kag-C1kT}.
Significantly, none of the three moves toward the value $1.0$ as $T\rightarrow 0$.  Rather,
they move towards a value close to 1.38 at zero temperature.  This indicates that rather than
becoming oriented with $\cos^{-1}(-\tfrac{1}{2})=120^{\circ}$ angular deviations as in a ground 
state, the neighboring dipoles tend to have a different average relative orientation.  That is, 
they may be considered to have an average near neighbor deviation closer to 
$\cos^{-1}(-\tfrac{1.38}{2})\approx 134^{\circ}$.  That would be an extra misalignment of neighbors
(towards being more anti-aligned), at the expense of extra uniaxial anisotropy energy (due to 
$K_1$ energy not being minimized).  Also note, all the correlations $C_{ab}(T), C_{bc}(T)$ and
$C_{ac}(T)$ behave the same as $C_1(T)$, but with half the amplitude.

\begin{figure}
\includegraphics[width=\figwidth,angle=-90.]{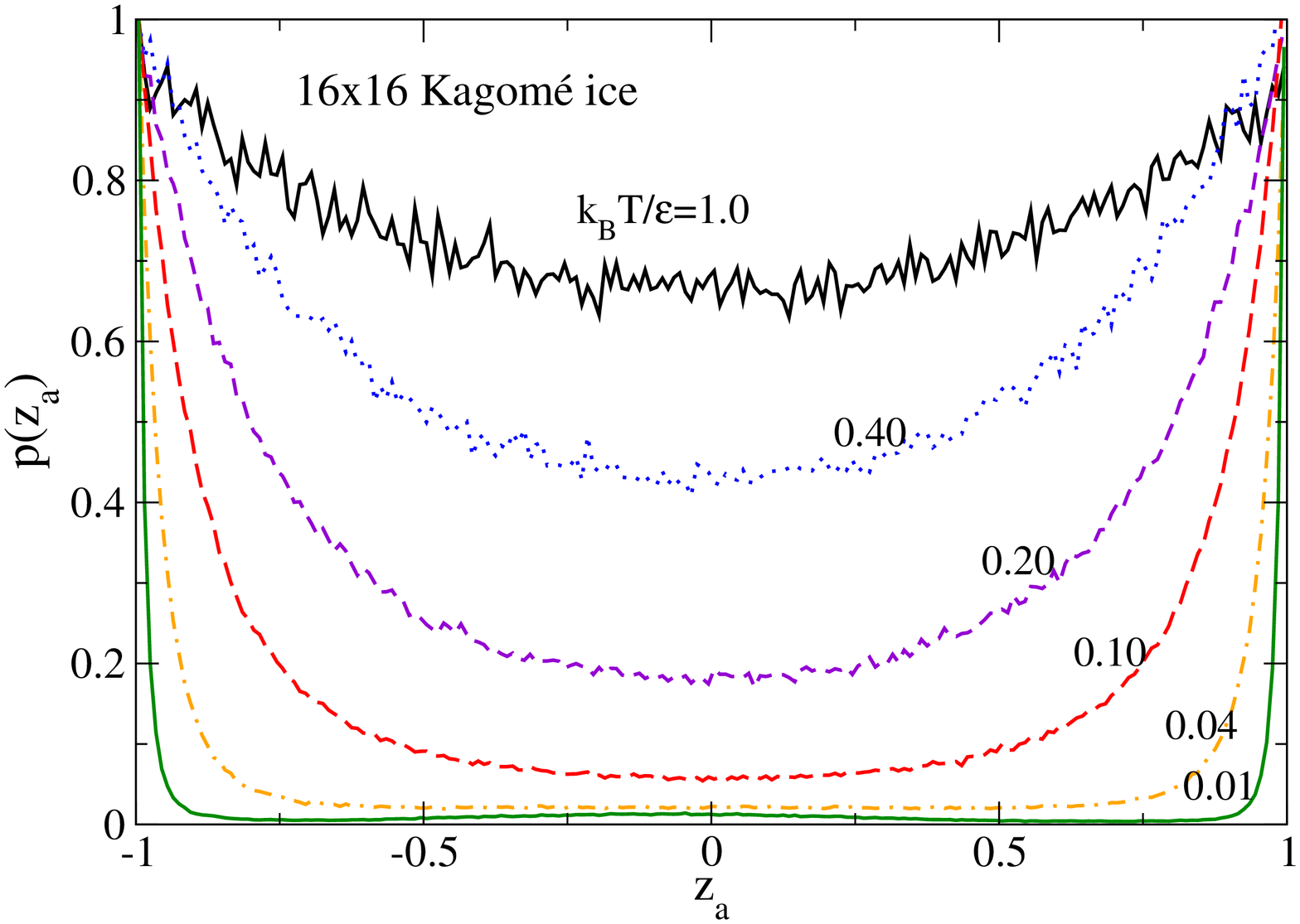}
\includegraphics[width=\figwidth,angle=-90.]{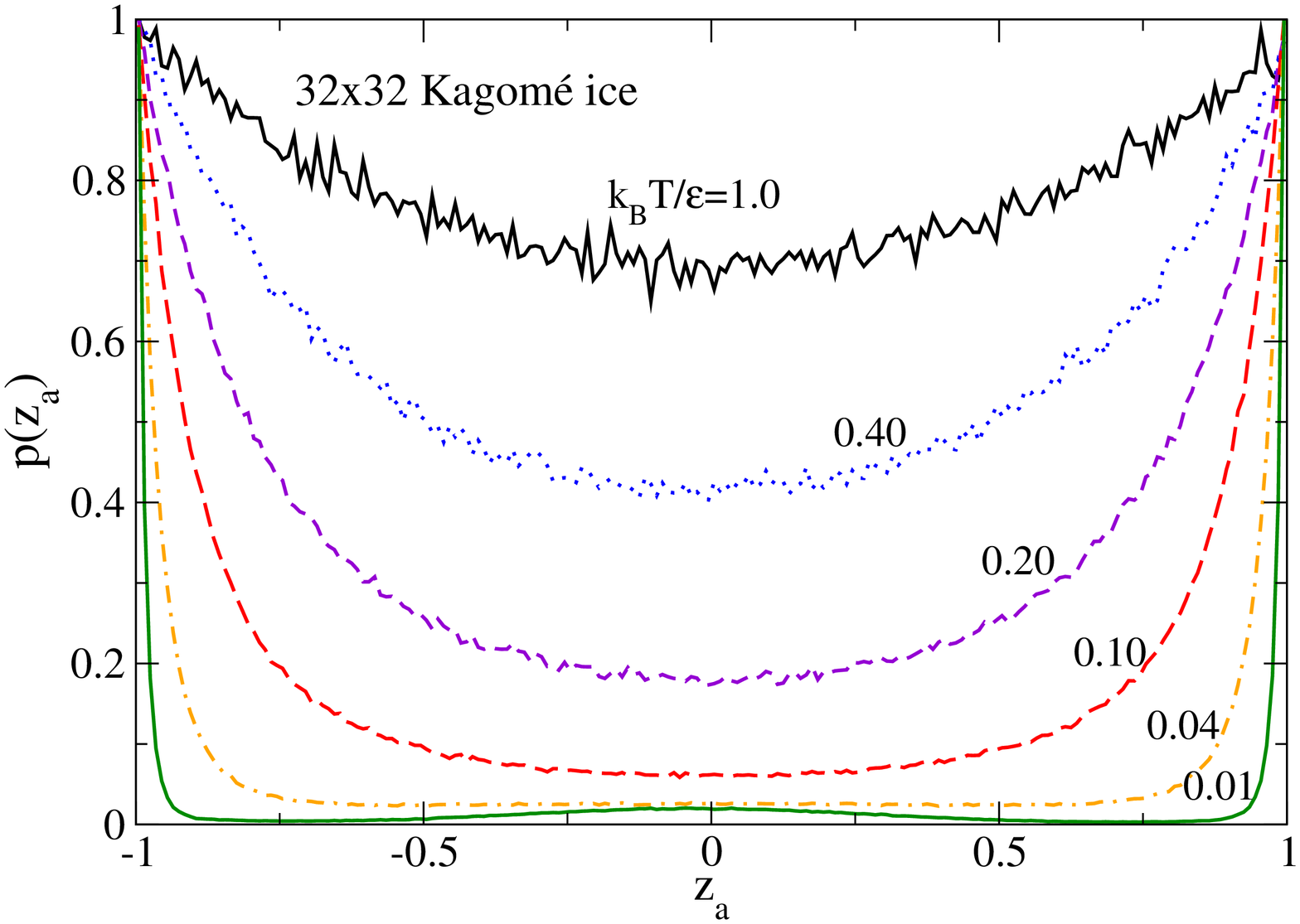}
\caption{\label{kag-pza} The raw probability distribution of a-sublattice order $z_a$, 
Eq.\ (\ref{localza}), for Kagom\'e spin ice in (a) a $16 \times 16$ system and (b) 
a $32 \times 32$ system.  Lables on curves give the dimensionless temperature.
The distribution changes from wide at high temperature (curvature controlled
by the uniaxial anisotropy $K_1$ relative to thermal energy) to being split into 
two llimiting ranges as the temperature is lowered. Very similar distributions 
hold for $z_b$ and $z_c$ on the other sublattices.}
\end{figure}

\begin{figure}
\includegraphics[width=\figwidth,angle=-90.]{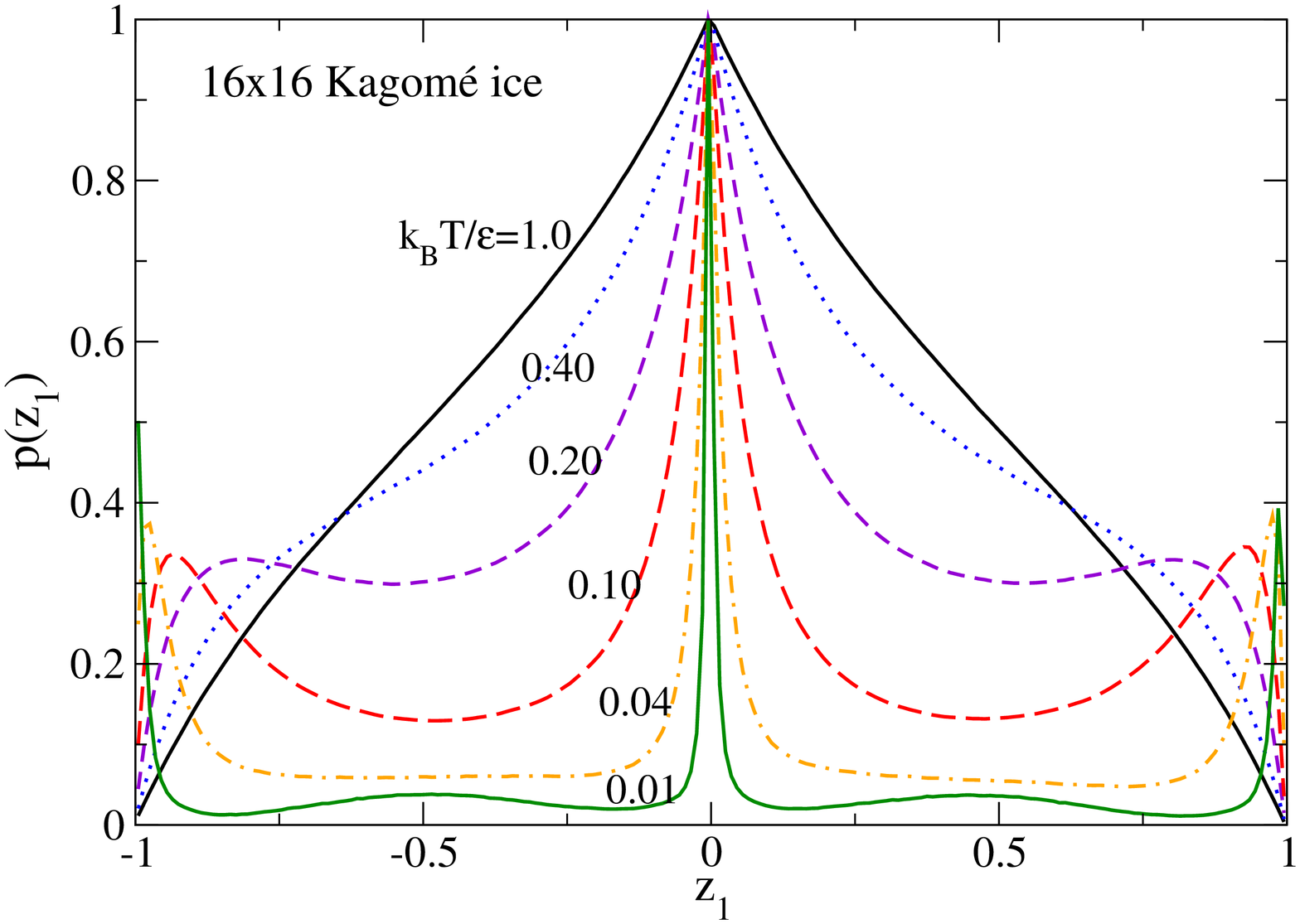}
\includegraphics[width=\figwidth,angle=-90.]{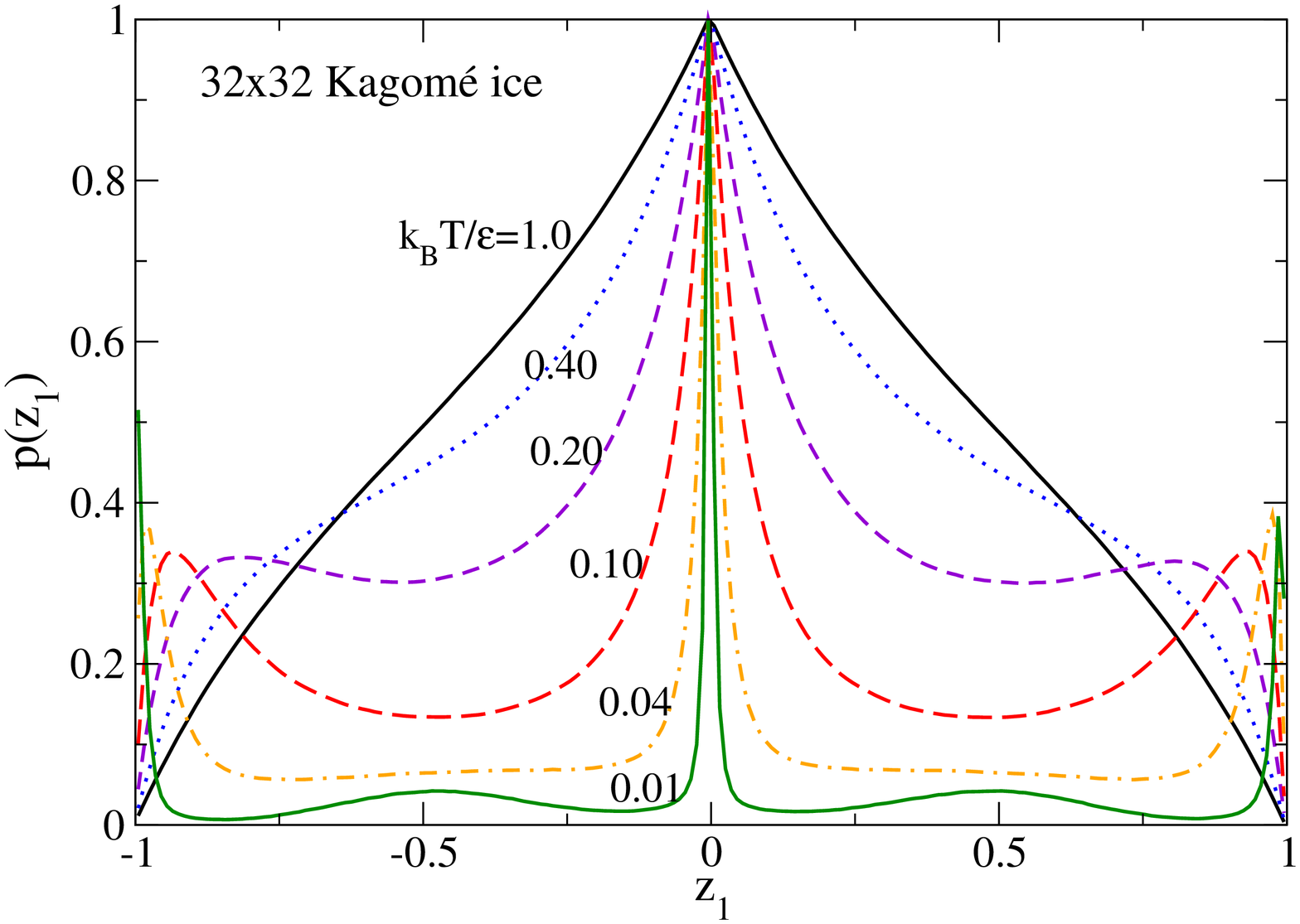}
\caption{\label{kag-pz1} Examples of the derived probability distribution of the local order
parameter $z_1$, see Eq.\ (\ref{z123}) and Eq.\ (\ref{localz1}), for Kagom\'e lattice spin ice, in 
(a) a $16 \times 16$ system and (b) a $32 \times 32$ system.  Note that it is derived from 
distributions of sublattice variables $z_b$ and $z_c$.  Different curves are labeled by the 
dimensionless temperature.  The distribution changes from wide at high $T$ (a triangular shape
results from uniform $p_b(z_b)$ and $p_c(z_c)$ in extreme high-$T$ limit) to being split into 
three branches at low $T$ ( $z_1$ near $0, \pm 1$). There is no relaxation into a pure
ground state.}
\end{figure}

Further indications of the local dipolar order can be seen in the probability distributions
for local order parameters on magnetic sublattices, $z_a, z_b, z_c$ and the derived quantities
for proximity to the ground states, $z_1, z_2, z_3$, see Eq.\ (\ref{z123}).  Consider the
probability distributions at different temperatures found for $p_a(z_a)$, shown in Fig.\
\ref{kag-pza}.  At high temperature the wide parabolic form resembles the distribution of
$z_a$ as found in square ice, Fig.\ \ref{sqr-pza}.  Recall that this parabolic form is
mostly due to a Boltzmann factor like $\exp\{K_1 z_a^2/k_B T\}$, due to the competition
of the uniaxial energy with the temperature.  As the temperature is lowered, however, there
is not a collapse of the distribution to one side or the other, as appeared in square ice.  
Instead,  the limiting states $z_a=-1$ and $z_a=+1$ remain nearly equally probable, even 
at very low temperature.  The distribution of $z_a$ (and also $z_b$ and $z_c$, not shown)
remains nearly symmetric.   The system does not move into a ground state, even in a local
sense.  This clearly is a manisfestation of the geometric frustration.  

The distribution of $z_1$ is shown in Fig.\ \ref{kag-pz1}, again for a set of temperatures.
The quantities $z_1, z_2, z_3$ give measures of the local proximity of the dipoles to a
ground state configuration.  The quantity $z_1$ for an individual dipole would take the value 
$z_1=+1$, for instance, together with $z_2=z_3=0$, if a dipole were oriented as in the ground 
state $\Psi_{\rm gs}^{1+}$.   Of course, the randomness in the system leads to a wide distribution
of $z_1$, and similarly, of $z_2$ and $z_3$ for projections into the other ground states.
At high temperatures, the distribution of $z_1$ shown in Fig.\ \ref{kag-pz1} is close to 
triangular.  A perfect triangle would result from Eq.\ \ref{localz1} if the distributions of 
$z_b$ and $z_c$ were perfectly uniform from $-1$ to $+1$.  Because there is a slight
curvature in $p_a(z_a)$, and similarly in $p_b(z_b)$ and $p_c(z_c)$, the distribution for
$z_1$ at high temperature becomes a slightly curved triangular shape.  This distribution
peaks at $z_1=0$ and goes to zero weight at $z_1 \rightarrow \pm 1$.  The form indicates
that the system is far from a ground state.  For lower temperatures, the distribution of
$z_1$ acquires a strong narrow peak around $z_1=0$ and peaks about half as strong at
$z_1=\pm 1$.  Note that the corresponding curves for $p_2(z_2)$ and $p_3(z_3)$ are essentially
the same as that for $p_1(z_1)$. The system does not condense into a particular ground state, 
which would have been indicated by having only one peak in each distribution.  The low-temperature 
distrubtion is indicative that the dipoles are about equally likely to be in any of the six possible
ground states.  

\begin{figure}
\includegraphics[width=\figwidth,angle=-90.]{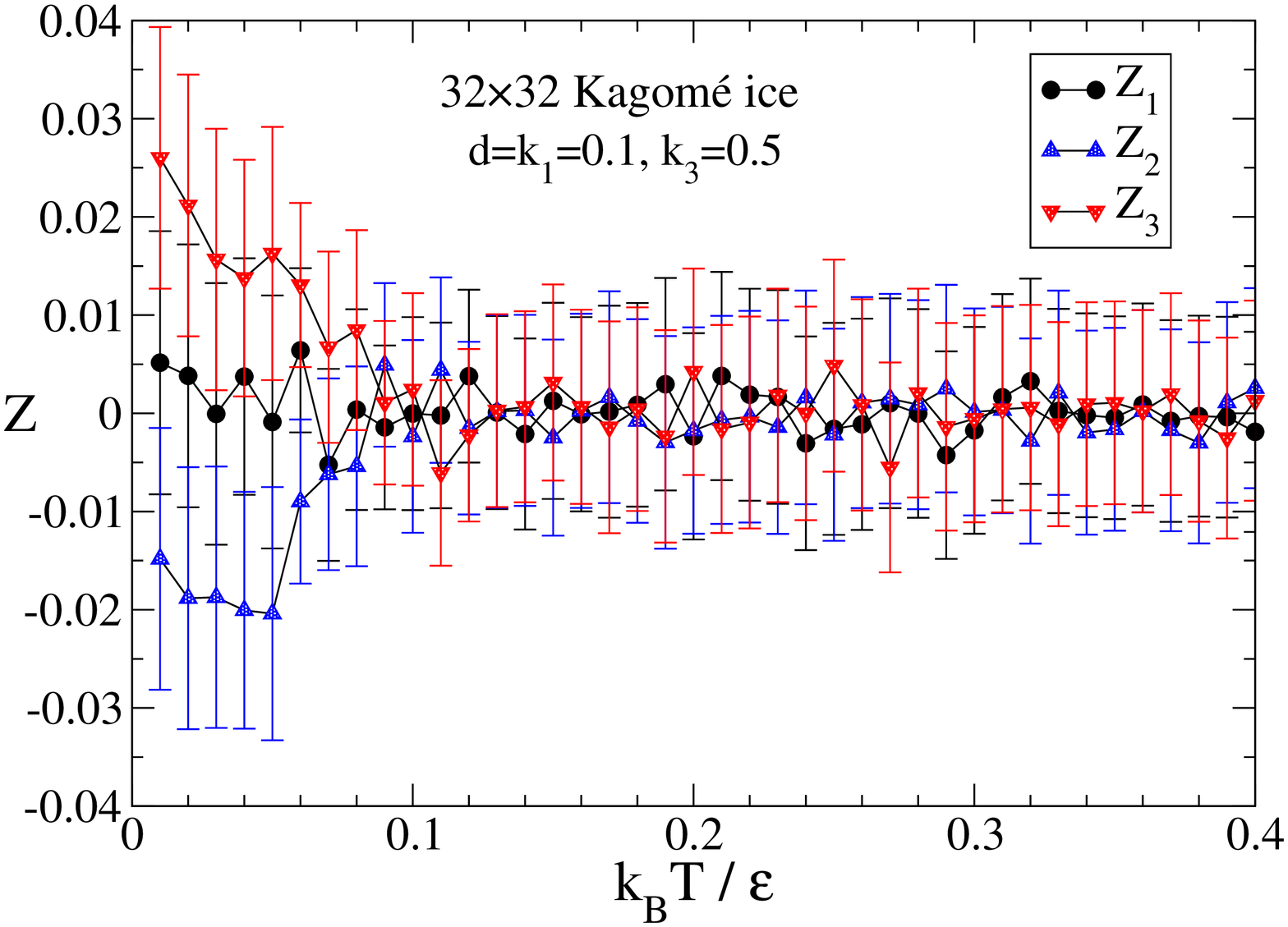}
\caption{\label{kag-ZkT} System order parameters for ground state projections,  for 
$32 \times 32$ Kagom\'e lattice spin ice, as defined in Eq.\ (\ref{Z123}).  
At high temperatures there is no order and $Z_1=Z_2=Z_3=0$.  At lower temperatures
only a very slight global ordering of the system is taking place.  If the system condensed 
into the $\Psi_{\rm gs}^{1+}$ ground states, for instance, one would obtain limiting 
values $Z_1=1, Z_2=Z_3=0$.}
\end{figure}

The order parameters $Z_1, Z_2, Z_3$ measure the tendency of the global system to be
ordered as one of the ground states, see the definition Eq.\ (\ref{Z123}).  Their
behavior with temperature is indicated in Fig.\ \ref{kag-ZkT}, for the $32 \times 32$
system.  For high temperatures, there is no particular ordering close to any ground
state, and all three parameters are zero.  If the system were to move into the 
$\Psi_{\rm gs}^{1+}$ ground state, for instance, the parameters take on the values
$Z_1=1, Z_2=Z_3=0$.  But as the temperature is lowered, there are only very slight
deviations in these parameters away from zero, hardly significant relative to the
error bars.  This is totally consistent with the above results for the distributions
of the local order parameters $z_1, z_2, z_3$.  Overall, the geometrical frustration
significantly prevents movement of the system into any region of phase space near
one of the ground states, when bringing the temperature from  higher to lower values.

\section{Discussion and Conclusions}

These studies  consider the spin ice islands as dipoles of fixed magnitude that
change directions in response to local uniaxial anisotropy ($K_1$), a planar anisotropy
($K_3$), and the long range dipolar interaction ($D$).   The thermodynamic properties 
within this approximation have been investigated by studying the Langevin dynamics of
the system.  

For square ice, at very high temperature, the charge densities for singly-charged and 
doubly-charged monopoles found in simulations go toward the expected asymptotic values, 
$\rho_1 \rightarrow \tfrac{1}{2}$, $\rho_2 \rightarrow \tfrac{1}{4}$, and total 
density $\rho \rightarrow \tfrac{3}{4}$.
Cooling the system from higher towards lower temperature results in
the system approaching one of the ground states, by a random choice between the two.
The evidence for this comes from different measured parameters, including the monopole
charge densities $\rho_1$ and $\rho_2$, the order parameter $Z$, the near neighbor 
correlation $N_{ab}$, and the probability distribution of local order parameters 
$z_a, z_b$ and $z$.  The order parameter approaches $Z=\pm 1$, the correlations tend 
to the value $N_{ab}=1$, and the probability distribution for $z$ becomes asymmetric 
with a peak near either $z=-1$ or $z=+1$.  The system smoothly moves towards one or 
the other ground state as $T \rightarrow 0$ without any strong frustration.  

For Kagom\'e ice, at very high temperature, the charge densities in simulations
go to their expected asymptotic values, $\rho_1 \rightarrow \tfrac{3}{8}$ for single
charges, and $\rho_3 \rightarrow \tfrac{3}{8}$ for triple charges, with total density
$\rho \rightarrow \tfrac{3}{4}$.  The total asymptotic density is the same as in
square ice.  As functions of increasing temperature, triple charges are generated at 
the expense of single charges.  Conversely, as the temperature is lowered, the Kagom\'e
system does not move easily to one of its six ground states, not even in any limited 
(local) sense.  We have characterized this frustrated dynamics by various measures,
including the near neighbor correlations between magnetic sublattices $C_{ab}, C_{bc}, C_{ac}$,
the correlations related to ground state order, $C_1, C_2, C_3$, the order parameters
on sublattices $Z_a, Z_b, Z_c$ and their counterparts $Z_1, Z_2, Z_3$, and also the
probability distributions for local order parameters $z_a, z_b, z_c$ and $z_1, z_2, z_3$.
Surprisingly, the zero temperature limit for the correlations is $C_1=C_2=C_3  \rightarrow 1.38$,
with $C_{ab}, C_{bc}$ and $C_{ac}$ going to half this value.  These do not present the
values expected of a ground state.  Similarly, the probability distributions for
$z_1, z_2, z_3$ have three peaks in the limit of low temperature,  and none of the order
parameters $Z_1, Z_2, Z_3$ become anywhere close to $\pm 1$,  showing that the system
remains far from a ground state.  Clearly with the set of six independent ground states
and rotational symmetry of the system, such freedom in phase space makes it nearly
impossible to relax directly to one of the ground states over any large system.

\appendix

\begin{figure}
\includegraphics[width=\figwidth,angle=0.]{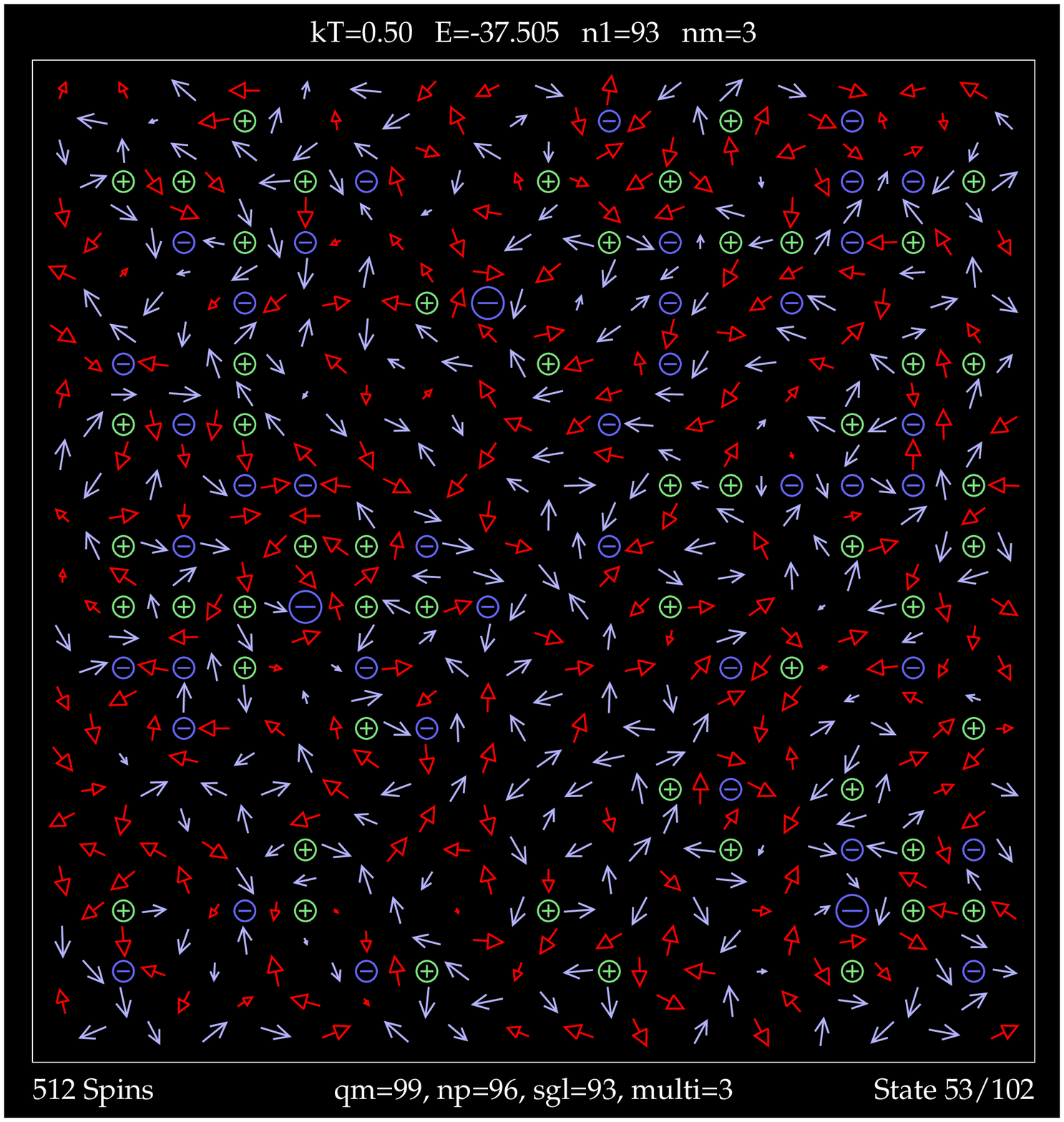}
\includegraphics[width=\figwidth,angle=0.]{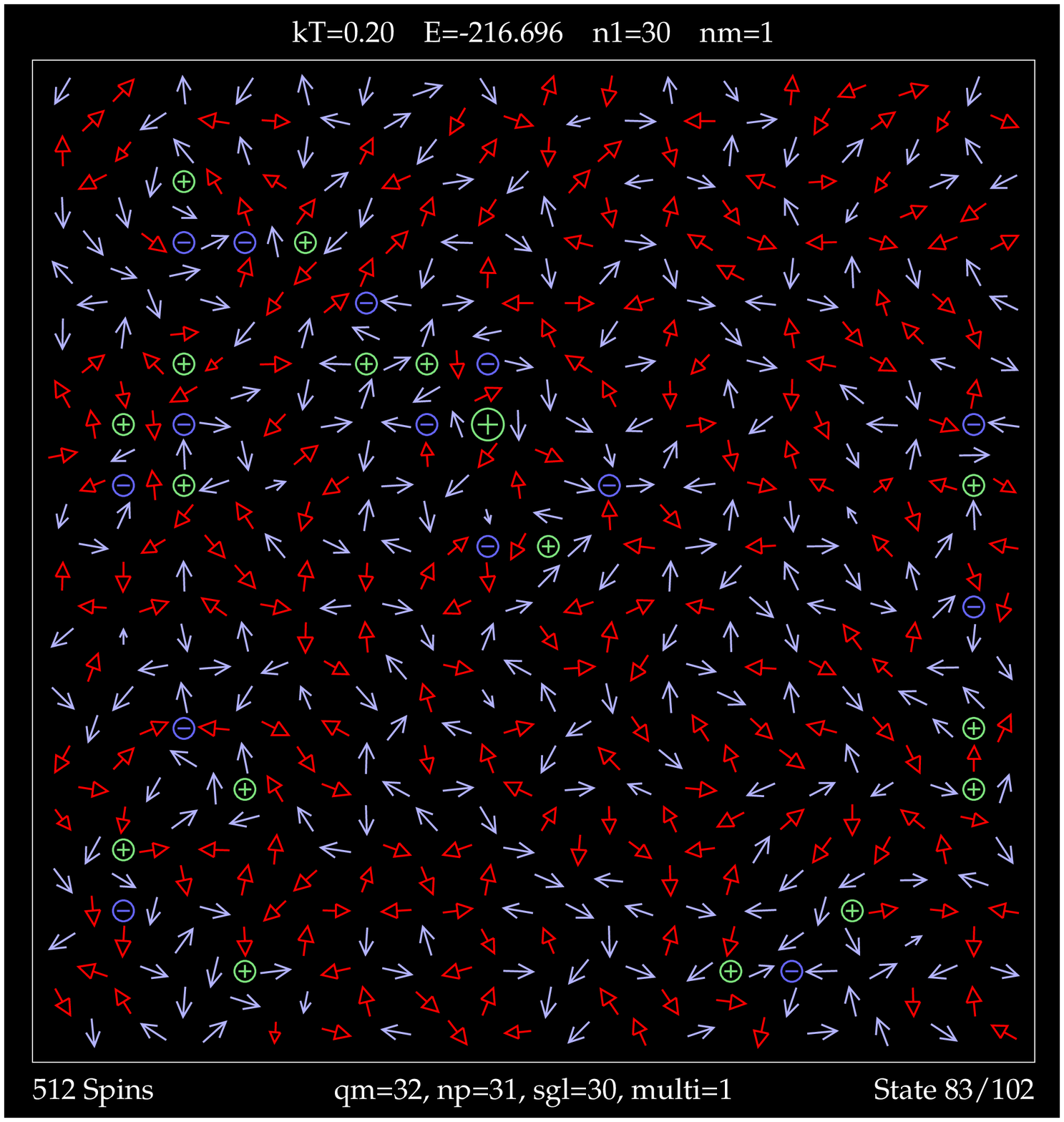}
\includegraphics[width=\figwidth,angle=0.]{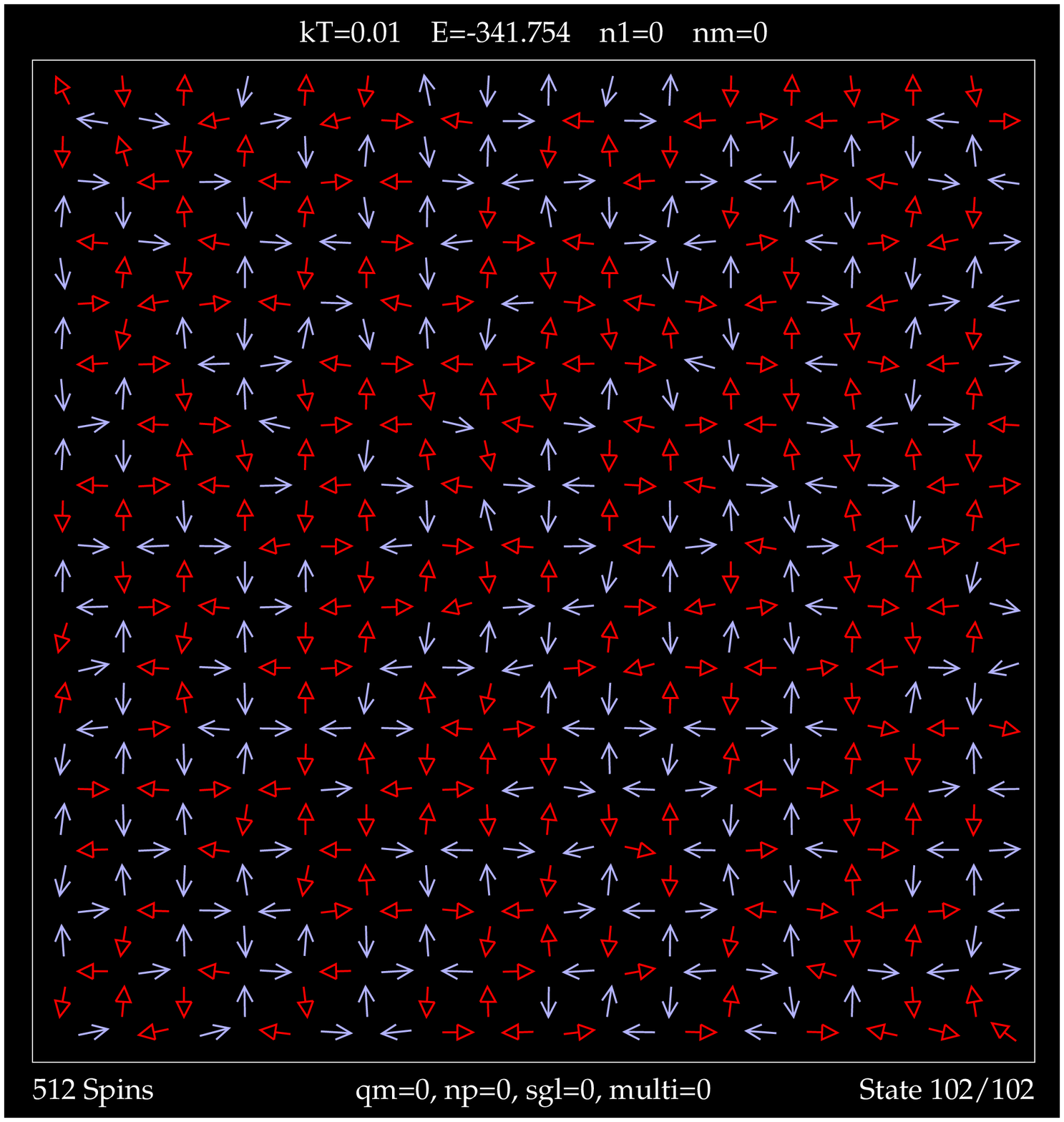}

\caption{\label{sqr-configs} Dipole configurations in $16\times 16$ square ice,
for temperatures (a) ${\cal T} = 0.50$,  (b) ${\cal T} = 0.20$, (c) ${\cal T} = 0.01$ .
Blue (red) indicates $+$ ($-$) $\mu_z$ components.
Small circles indicate monopole charges, larger circles are double charges.}
\end{figure}

\begin{figure}
\includegraphics[width=\figwidth,angle=0.]{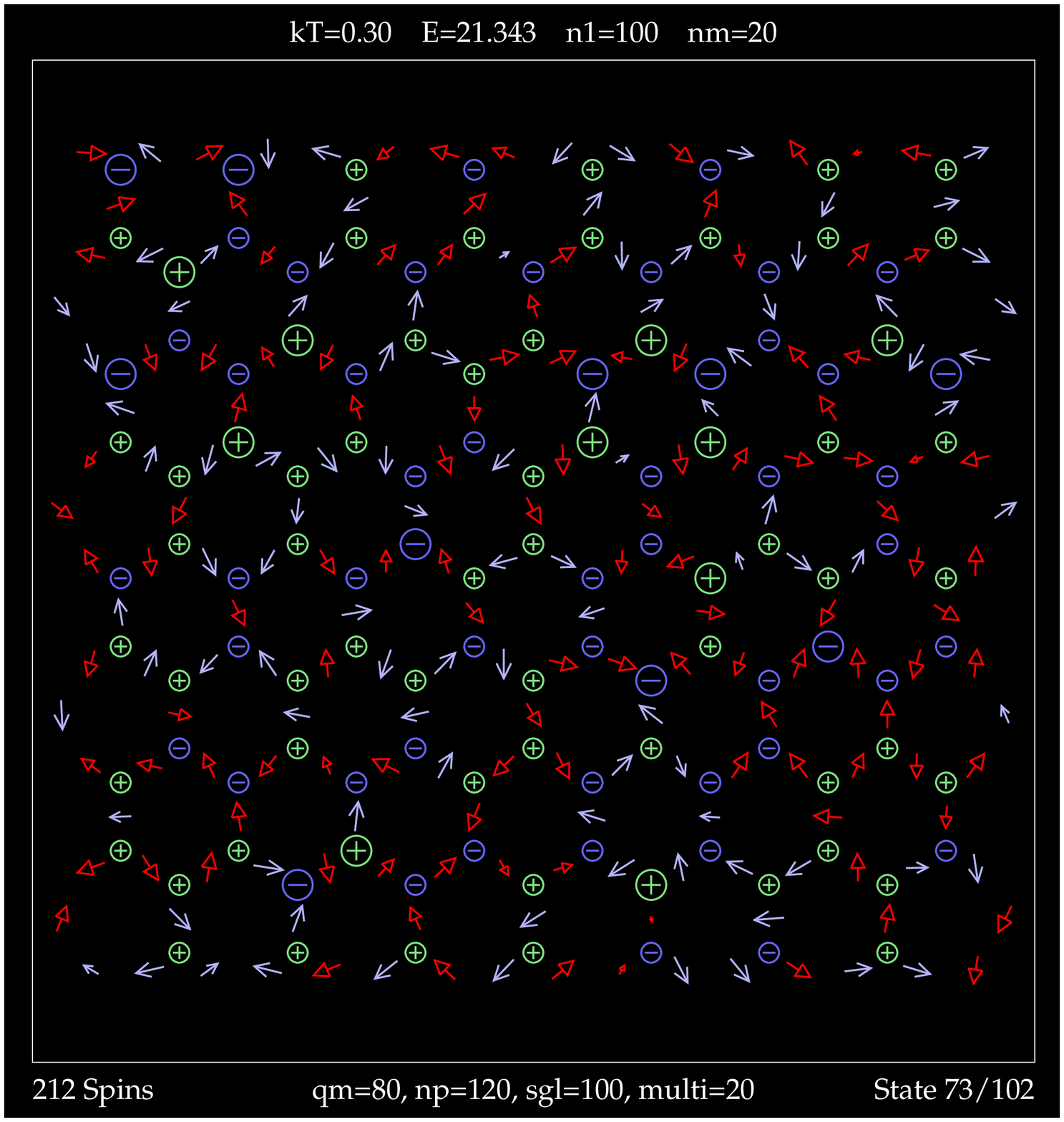}
\includegraphics[width=\figwidth,angle=0.]{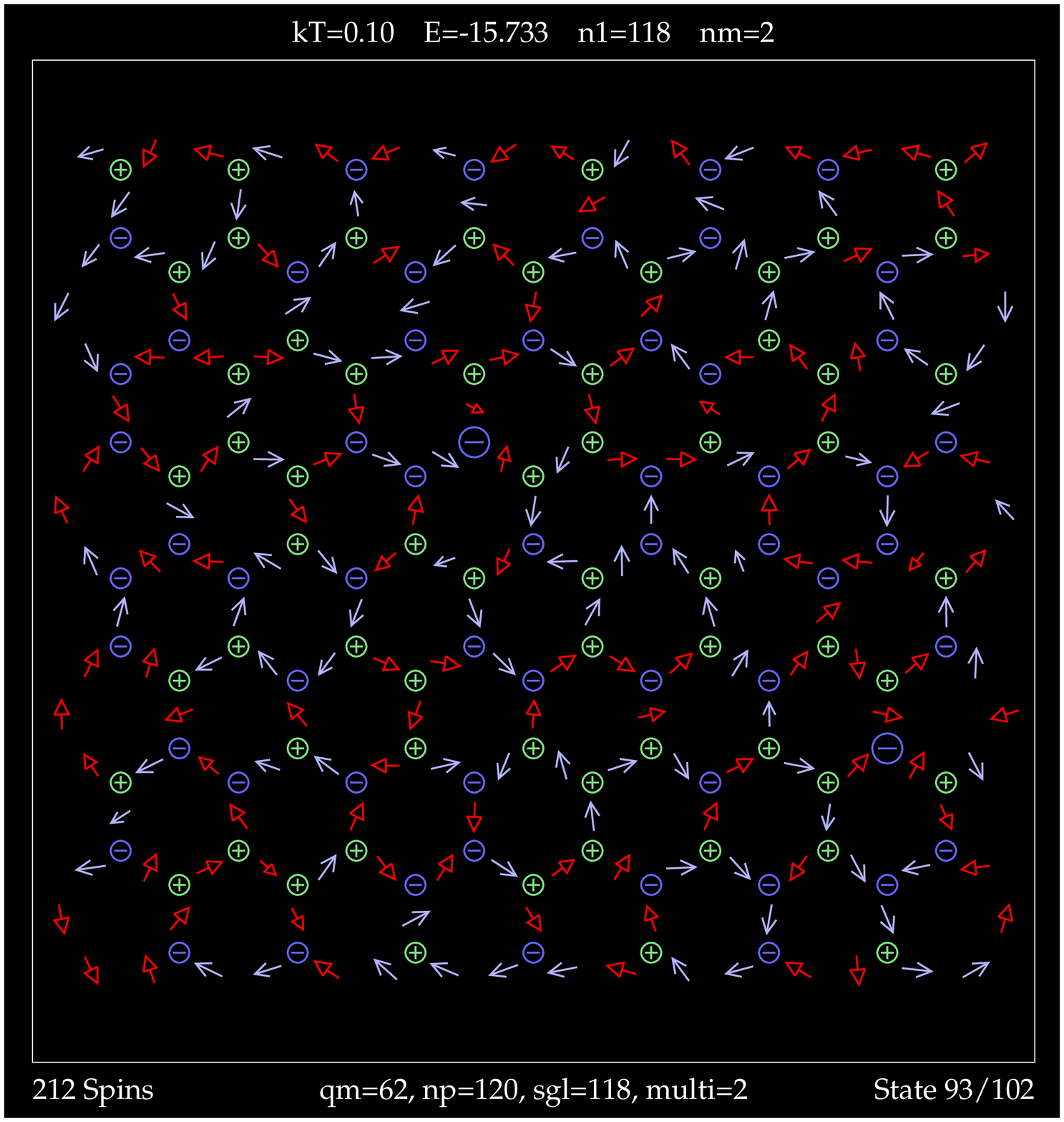}
\includegraphics[width=\figwidth,angle=0.]{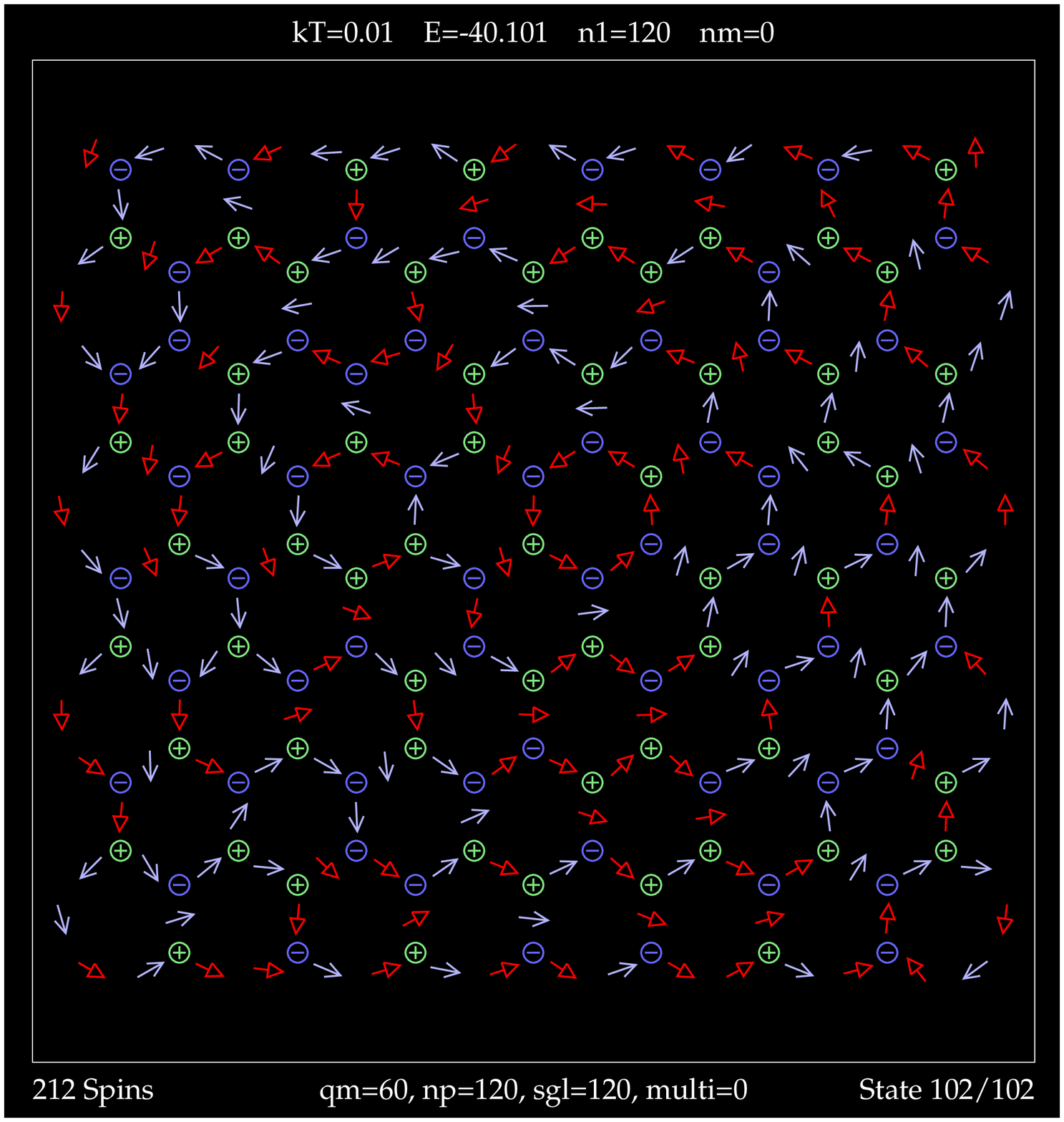}

\caption{\label{kag-configs} Dipole configurations in $16\times 16$ Kagom\'e ice,
for temperatures (a) ${\cal T} = 0.30$,  (b) ${\cal T} = 0.10$, (c) ${\cal T} = 0.01$.
Blue (red) indicates $+$ ($-$) $\mu_z$ components.  Small circles indicate single charges 
$q=\pm \tfrac{1}{2}$, larger circles are triple charges $q=\pm \tfrac{3}{2}$.}
\end{figure}

\section{Typical dipolar states}
An overview of the dipolar configuration  in $16 \times 16$ square ice is shown
in Fig.\ \ref{sqr-configs}, for dimensionless temperatures ${\cal T} = 0.50$ 
(above the transition), ${\cal T} = 0.20$ (just below the transition) and
${\cal T} = 0.01$ (nearly in the ground state).  Note that at ${\cal T} = 0.01$,
the dipoles alternate direction along any row or column, as expected in one
of the ground states.  At the higher temperatures, circled plus and minus signs
indicate discrete monopole charges.  The larger circles are the double charges.

For Kagom\'e ice, similar plots are shown in Fig.\ \ref{kag-configs} for ${\cal T} = 0.30$ 
(well above the transition), ${\cal T} = 0.10$ (barely above the transition) and 
${\cal T} = 0.01$ (very low temperature).  Smaller circled plus or minus signs are
single charges ($q=\pm \tfrac{1}{2}$) and larger circles are triple charges 
($q=\pm \tfrac{3}{2}$).  At ${\cal T} = 0.01$, the system is totally filled with 
single charges, however, in a ground state they would alternate in sign between every 
bond of the lattice.\\


Acknowledgements\\

The authors would like to thank the Brazilian agencies CNPq, FAPEMIG and CAPES. WAMM and ARP also thank the hospitality of Max Planck Institute for the Physics of Complex Systems in Dresden, Germany. 

\end{document}